\documentclass{sig-alternate}

\usepackage[final]{listings}
\usepackage{color}
\usepackage{mathptmx}
\usepackage{graphicx}
\usepackage{color}
\usepackage{amsmath}
\usepackage{amsfonts}
\usepackage{amssymb}
\usepackage{threeparttable}
\usepackage{booktabs}
\usepackage{array}
\usepackage{fixltx2e}

\graphicspath{{figures/}}

\newcommand{\rlm}{Roof\/line model}
\newcommand{\ecmm}{ECM model}

\newcommand{\bq}{\begin{equation}}
\newcommand{\eq}{\end{equation}}

\newcommand{\muops}{$\mu$ops}
\newcommand{\flops}{\mbox{flops}}

\newcommand{\FS}{\mbox{flop/s}}
\newcommand{\GBS}{\mbox{GB/s}}

\newcommand{\GFS}{\mbox{Gflop/s}}

\newcommand{\LUPS}{\mbox{LUP/s}}
\newcommand{\MLUPS}{\mbox{MLUP/s}}

\newcommand{\GHZ}{\mbox{GHz}}

\newcommand{\LUP}{\mbox{LUP}}

\newcommand{\bytes}{\mbox{B}}
\newcommand{\byte}{\mbox{B}}
\newcommand{\bit}{\mbox{bit}}
\newcommand{\bits}{\mbox{bits}}

\newcommand{\cycles}{\mbox{cy}}
\newcommand{\eos}{~.}

\newcommand{\olsep}{\|}
\newcommand{\nolsep}{|}
\newcommand{\ecmspace}{\,}
\newcommand{\ecm}[6]{\mbox{$\left\{{#1}\ecmspace\olsep\ecmspace {#2}\ecmspace\nolsep\ecmspace {#3}\ecmspace\nolsep\ecmspace {#4}\ecmspace\nolsep\ecmspace {#5}\right\}\ecmspace{#6}$}}
\newcommand{\epsep}{\rceil}
\newcommand{\ecmp}[5]{\mbox{$\left\{{#1}\ecmspace\epsep\ecmspace {#2}\ecmspace\epsep\ecmspace {#3}\ecmspace\epsep\ecmspace {#4}\right\}\ecmspace{#5}$}}
\newcommand{\ecme}[5]{\mbox{${#1}\ecmspace\epsep\ecmspace {#2}\ecmspace\epsep\ecmspace {#3}\ecmspace\epsep\ecmspace {#4}\ecmspace{#5}$}}

\begin{document}
\conferenceinfo{ICS}{'15 Newport Beach, California USA}
\title{Quantifying performance bottlenecks of stencil computations using the Execution-Cache-Memory model}
\numberofauthors{4}
\author{
\mbox{Holger~Stengel\qquad Jan~Treibig\qquad Georg~Hager\qquad Gerhard~Wellein}\\
\and
\affaddr{Erlangen Regional Computing Center (RRZE)}\\
\affaddr{Friedrich-Alexander University of Erlangen-Nuremberg}\\
\affaddr{Erlangen, Germany}\\
\email{\{holger.stengel,jan.treibig,georg.hager,gerhard.wellein\}@fau.de}
}

\emergencystretch5pt

\newcommand{\comment}[1]{\color{red}!!! {#1} !!!\color{black}}


\maketitle

\begin{abstract}
Stencil algorithms on regular lattices appear in many fields of computational
science, and much effort has been put into optimized implementations. Such
activities are usually not guided by performance models that provide estimates
of expected speedup. Understanding the performance properties and bottlenecks
by performance modeling enables a clear view on promising optimization
opportunities. In this work we refine the recently developed
Execution-Cache-Memory (ECM) model and use it to quantify the performance
bottlenecks of stencil algorithms on a contemporary Intel processor. This
includes applying the model to arrive at single-core performance and
scalability predictions for typical ``corner case'' stencil loop kernels.
Guided by the \ecmm\ we accurately quantify the significance of ``layer
conditions,'' which are required to estimate the data traffic
through the memory hierarchy, and study the impact of typical optimization
approaches such as spatial blocking, strength reduction, and temporal blocking 
for their expected benefits. We also compare the \ecmm\ to the widely known
\rlm.
\end{abstract}


\section{Introduction and related work}
\label{sec:intro}
Stencil computations on regular lattices are important in many
computational science applications operating on regular lattices. The
combination of low computational intensity and their iterative nature
have made them a popular target for loop optimizations.  Beyond
established cache blocking techniques, leveraging SIMD capabilities
\cite{Henretty:2011} and an efficient shared memory parallelization
\cite{th09} are subject to recent research. An overview about the
state of the art in stencil optimizations can be found in a review by
Datta et al.~\cite{datta09}.

In the compiler community modeling approaches based on cache usage have been
employed for stencil optimizations more than a decade ago \cite{Leopold:2002}. 
Modeling was also used to estimate the optimal performance of stencil codes
and to validate the effectiveness of applied optimizations. One
can roughly separate those efforts into two categories: ``black box''
modeling, which relies on statistical methods and machine learning to describe
and predict performance behavior based on observed data such as hardware
performance metrics \cite{Rahman:2011}, and ``white box'' modeling, 
which uses simplified machine models to describe the interaction of code with the
hardware from ``first principles.'' Simple bottleneck-centric
approaches based on memory bandwidth vs.\ computational peak performance
 have been in use for a long
time~\cite{Callahan88,hockney89}.
This line of thinking was popularized and refined by Williams et al.\
\cite{Williams:2009}, and is now known as the ``\rlm.'' The \rlm\ is
tremendously useful in getting first estimates of the ``light speed'' of a loop
on a given architecture. Its central assumption is that data transfers
through the memory hierarchy overlap with code execution on the core(s).
This implies that
data access latencies are assumed to be hidden by prefetching
mechanisms.
However, the \rlm\ cannot describe relevant bottlenecks
beyond memory bandwidth and peak performance. Especially in long-range stencils common
in seismic imaging there is no single dominating bottleneck anymore and the
influence of in-cache transfers on performance is significant. The recently
introduced \ecmm\ provides an accurate single-core performance and scaling prediction. 
It can be seen as an extension and refinement
of the \rlm\ for multicore processors. Like the \rlm\ it focuses on resource
utilization but considers contributions from the complete memory
hierarchy. The \ecmm\ was first introduced
in \cite{th09} and later refined and combined with a multicore power model in
\cite{hager:cpe}. It provides clear insights into the relevant
contributions of the hardware model to the performance 
of a given loop code and therefore allows a clear identification of
optimization opportunities. Based on the single-core performance estimate
it also predicts a performance saturation point.
This paper refines the \ecmm\ with clear rules for overlapping execution
and data transfers and introduces an accessible notation to ease the usage
and description of the model. The power of the model to validate and guide
performance engineering efforts is demonstrated on the example of three relevant
stencil code patterns.

The paper is structured as follows. In Sect.~\ref{sec:testbed} we describe the
architecture of the test machine used for measurements. Section~\ref{sec:ecm} 
introduces and refines the \ecmm\ on the example of simple
streaming loop kernels. In Sections \ref{sec:j2d}, \ref{sec:uxx}, and 
\ref{sec:longstencil} we apply the model to successively more complex stencils to show
how it can identify bottlenecks, reveal optimization opportunities, and prevent
misleading conclusions.
The paper concludes in Sect.~\ref{sec:conclusion}.
\section{Experimental Testbed}
\label{sec:testbed}
\begin{table}[tb]
\renewcommand{\arraystretch}{1.2}
	\centerline{\begin{tabular}{lccc}
        \hline
	    Microarchitecture                &Intel SandyBridge-EP \cite{intelopt}\\
	    Model                            &Xeon E5-2680\\
        \hline
	    Clock speed (fixed)           &2.7\,GHz\\
        Cores/Threads                    &8/16\\
	    SIMD Support                     &SSE (128\,bit), AVX (256\,bit)\\
        \hline
	    L1/L2/L3 Cache                   &8$\times$32\,kB/8$\times$256\,kB/20\,MB\\
	    Main Memory Configuration        &4 channels DDR3-1600\\
        Update Benchmark Bandwidth       &40\,GB/s\\
        \hline
	\end{tabular}\smallskip}
    \caption{Test machine specifications. All information is shown for one socket.
    ``Update'' is a multi-threaded streaming benchmark that modifies an array in memory.}
    \label{tab:arch}
\end{table}
Unless noted otherwise, a
standard two-socket server based on eight-core SandyBridge-EP (SNB) 
CPUs was used for the measurements. The core supports SSE and
AVX SIMD instruction set extensions. Using floating-point arithmetic, each core
can execute one multiply and one add instruction per cycle, leading to a peak
performance of eight double precision (DP) or 16 single precision (SP) \flops\
per cycle. Memory is connected via four DDR3-1600 memory channels per socket. 
The SNB core can
sustain one full-width AVX load and one half-width AVX store per cycle. With
SSE or scalar execution, these limits are changed: In both cases the core can
sustain either one load and one store, or two loads per cycle, to the effect
that many loops do not show a 4$\times$ speedup of core execution when going
from scalar mode to AVX. The L2 cache sustains refills and evicts to and
from L1 at 256\,\bits\ per cycle (half-duplex). A full 64-\byte\ cache line 
transfer thus takes two cycles. The L3 cache is segmented, with one segment
per core. All segments are connected by a ring bus. Each segment has the same
bandwidth capabilities as the L2 cache, i.e., it can sustain 256\,\bits\ per cycle
(half-duplex) to and from L2. This means that the L3 cache is
usually not a bandwidth bottleneck, which is an improvement compared to
previous Intel processors. An overview of the  specifications is
given in Table \ref{tab:arch}. In Sect.~\ref{sec:uxx} we use a
$3.0\,\GHZ$ Intel Ivy Bridge CPU for comparison. The relevant differences
will be described there.

Measurements for the vector summation example were performed with LIKWID-bench 
\cite{likwid-psti}, and the Intel C Compiler (version 13.1.3) was used to
compile the stencil codes. The clock frequency was fixed to 2.7\,\GHZ.

\section{The ECM performance model}
\label{sec:ecm}
The \ecmm\ delivers a prediction of the number of CPU cycles required
to execute a certain number of iterations $n_\mathrm{it}$ of a given
loop on a single core. Since the smallest amount of data that can be
transferred between adjacent cache levels is usually a cache line
(CL), $n_\mathrm{it}$ is typically chosen to be ``one cache line's
worth of work.'' With a 64-\byte\ CL, this amounts to
$n_\mathrm{it}=8$ for a streaming kernel with double precision
data and stride-one data access to all arrays (such as the STREAM
benchmarks). Depending on the data types and access characteristics
this number may vary, however.

\subsection{Model input, construction, and assumptions}

Some input data is required to construct the \ecmm\ for a loop code on
a given architecture. We consider two parts of the input separately:
The time to execute the instructions on the processor core, assuming
that there are no cache misses, and the time to transfer data between
its initial location and the L1 cache. For the sake of clarity we review and refine
the model, which was first introduced in \cite{th09,hager:cpe}, using
simple streaming loops such as DAXPY (\verb.A(:)=A(:)+s*B(:).) 
before turning to more complex cases.

\subsubsection{In-core execution time}
\label{sec:incore}

\begin{figure}[tb]
\includegraphics*[width=\linewidth]{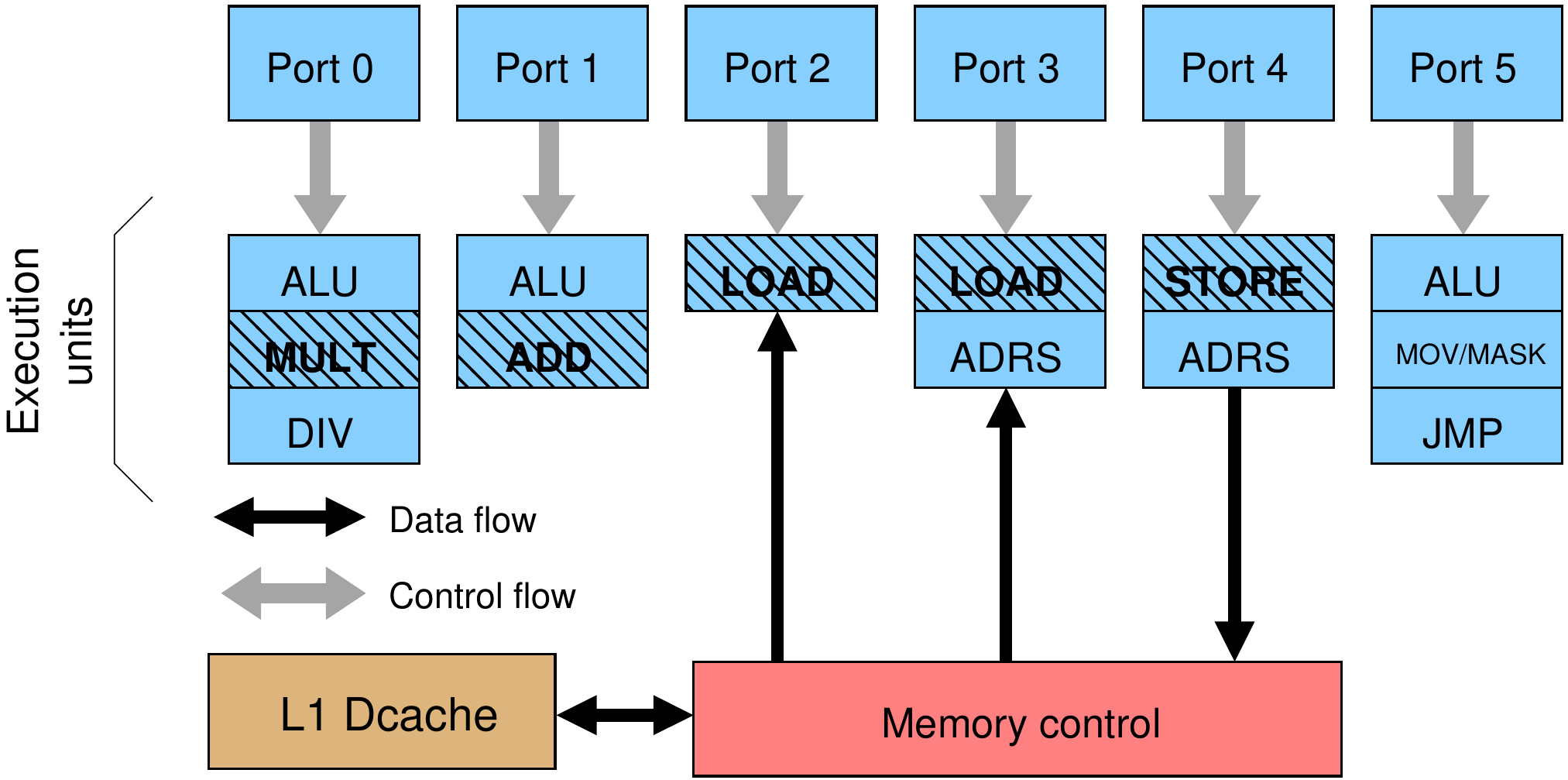}
\caption{\label{fig:gen_arch}Simplified view of execution resources on
 a modern out-of-order CPU core.}
\end{figure}
To determine the in-core execution time a simple model for instruction
throughput capabilities of a microarchitecture is required. On Intel processors
the port scheduler model can be used for this purpose.
Figure~\ref{fig:gen_arch} shows the port model for the Intel Sandy Bridge
microarchitecture. We assume that all instructions in a loop are independently
scheduled to the ports and executed out of program order within the
execution units $u_i$.
The execution time is then set by the unit that takes the
most cycles to execute the instructions:
\bq\label{eq:tcore}
T_\mathrm{core} = \max_iT_{u_i}\eos
\eq
An additional limitation that may be relevant is the overall maximum 
instruction throughput of four instructions
per cycle on Intel microarchitectures.\footnote{The term ``instruction'' is
used in a microarchitectural context; on Intel architectures these
are called ``\muops.''}
For example, the loop body of the DAXPY operation
\begin{lstlisting}
 for(i=0; i<N; ++i)
   a[i] = a[i] + s * b[i];
\end{lstlisting}
consists of two loads, one store, one add, and one multiply
instruction regardless of SIMD vectorization. In addition there
is typically one integer add, one integer compare, and one conditional
branch in the body to implement the ``loop mechanics.'' If the loop is
unrolled, the relative share of these instructions is smaller and can
usually be neglected. With AVX vectorization, one work unit (eight
iterations) comprises four load, two store, two add, and two multiply instructions,
which require four cycles to execute on the SNB core. In this
case the load (and store) pipeline throughput is the bottleneck. In
Fig.~\ref{fig:daxpy_serial} we visualize this situation in a timeline
diagram (cycles 1 to 4). Since there are no loop-carried dependencies in the code, the
four-cycle prediction holds if the loop is sufficiently unrolled.
In complex cases one can employ the \emph{Intel Architecture Code
 Analyzer} (IACA) \cite{iacaweb} to get an estimate of the loop body
execution time under the throughput assumption.

\subsubsection{Data transfers through the memory hierarchy}
\label{sec:ddelay}

Data not present in the L1 cache must be fetched from lower
levels of the memory hierarchy before it can be processed in CPU
registers, and modified data must be evicted to make room for new
cache lines. We call the time needed for these transfers the ``transfer
time.'' Again we assume the best possible case of perfect streaming,
i.e., there are no latency effects and the cost for transferring one
CL can be computed using the maximum bandwidth. For the
SNB architecture, one CL transfer takes two cycles between
adjacent cache levels. Getting a 64-\byte\ CL from memory to L3 or back
takes $64\,\bytes\cdot f/b_\mathrm S$ cycles, where $f$ is the
CPU clock speed and $b_\mathrm S$ is the achievable memory bandwidth
as measured by a suitable streaming benchmark.

Calculating the transfer times through the memory hierarchy requires
an accurate account of the data volumes between all cache levels. For the
DAXPY loop kernel discussed above this means that, due to the
cache hierarchy being inclusive, three CLs must be transferred between
each pair of adjacent memory levels: two for loading the data for
\verb.a. and \verb.b., and one for evicting modified
data of \verb.a.. This leads to transfer times of six cycles
each between cache levels, and 13 cycles between memory and L3 cache
(at a maximum socket bandwidth of 40\,\GBS\ and a clock frequency of
2.7\,\GHZ). If the ``uncore'' part of the chip, especially the L3 cache,
has a clock speed that is set independently from the core clock, a
correction factor can be applied to the L3-L2 cycle count.


\subsubsection{Single-core prediction and examples}

The in-core execution and transfer times described above must be put together
to arrive at a prediction of single-thread execution time. Therefore we must
specify which of the runtime contributions can overlap with each
other. If $T_\mathrm{data}$
is the transfer time (L1 downward, as described above),
$T_\mathrm{OL}$ is the part of the core execution that overlaps with
the transfer time, and $T_\mathrm{nOL}$ is the part that does not, then
\begin{eqnarray}
T_\mathrm{core} & = & \max\left(T_\mathrm{nOL},T_\mathrm{OL}\right)\\
T_\mathrm{ECM} & = & \max(T_\mathrm{nOL}+T_\mathrm{data},T_\mathrm{OL})\label{eq:T}\eos
\end{eqnarray}
There are two open questions here: (i) Which components of
$T_\mathrm{core}$ in (\ref{eq:tcore}) are overlapping? 
(ii) How is $T_\mathrm{data}$ composed of the different data transfer
contributions along the memory hierarchy?

The \ecmm\ makes the following fundamental assumptions for single-core
performance prediction: 
(1) All instructions in the loop are independently scheduled to the
  execution ports and executed out of program order within the
  execution units. The unit that takes the longest time executing its
  instructions is the bottleneck on the core level (see
  Sect.~\ref{sec:incore}). The maximum throughput capabilities of the microarchitecture
  are a further limitation.
(2) Loads (more specifically, cycles in which loads are retired) do
  not overlap with any other data transfer in the memory
  hierarchy. All other instructions, and also pipeline bubbles,
  overlap perfectly with data
  transfers. This means that only the time needed for the loads to the
  L1 cache contributes to $T_\mathrm{nOL}$ (store-bound loop kernels
  show significant overlap).
(3) The transfer times up to the L1 cache are mutually non-overlapping,
  i.e., all cache line transfers across the memory hierarchy are 
  serialized.

The non-overlapping assumptions seem very pessimistic; they were
motivated by the observation that the L1 cache on Intel processors can
either communicate with the L2 cache \emph{or} the registers, but not
both in the same cycle. We will show below that they lead to the most
accurate description of performance behavior.
Due to this non-overlap assumption the \ecmm\ does not provide an 
absolute upper performance bound. In contrast to the \rlm\ it allows
for performance measurements to exceed
the model prediction.

\begin{figure}[tb]
\centerline{\includegraphics*[width=0.9\linewidth]{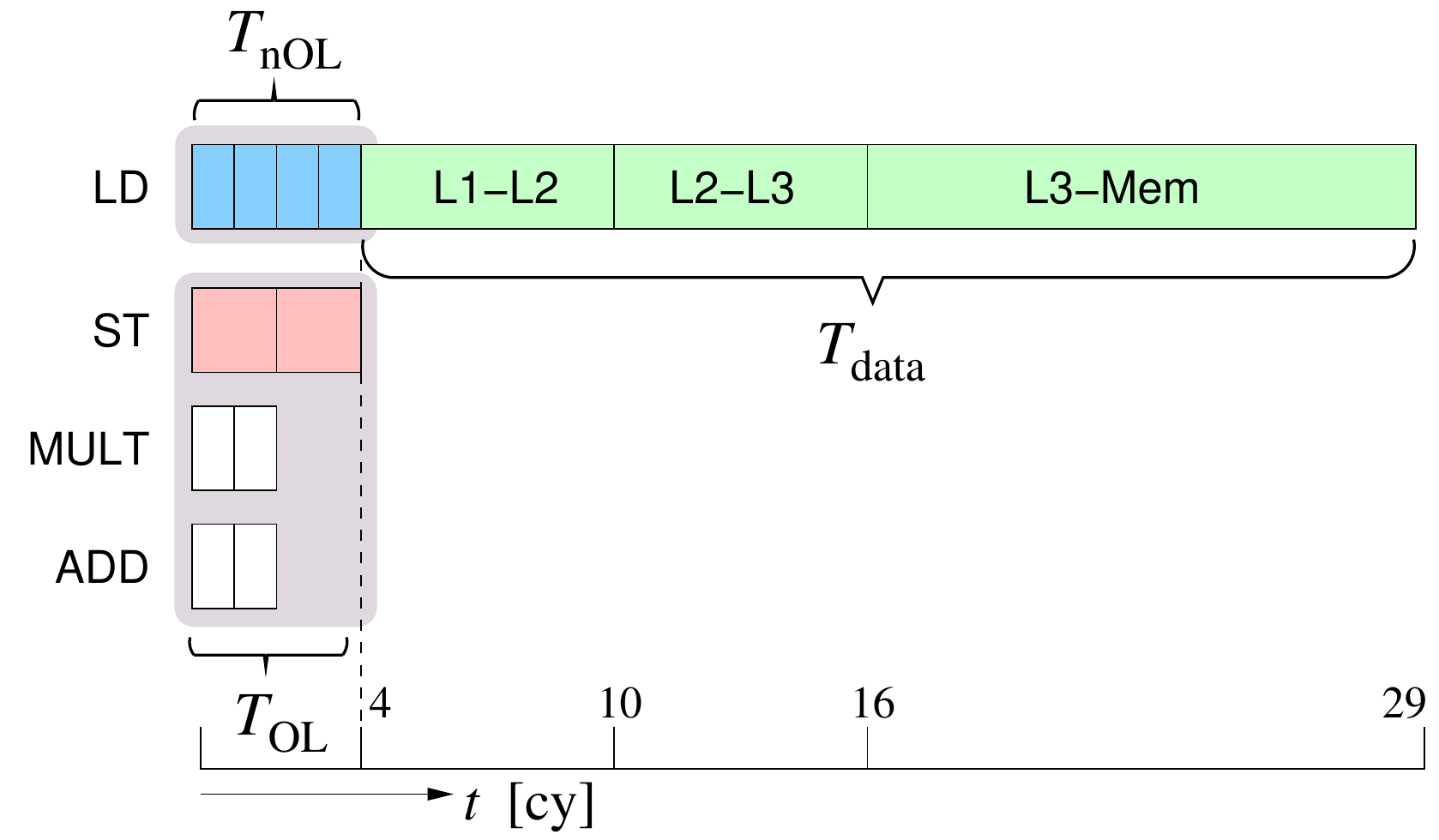}}
\caption{\label{fig:daxpy_serial}Single-core \ecmm\ for the DAXPY
 loop kernel on Intel SNB. The predicted number of cycles per CL
 is 29 if the data is in memory, 16 for the L3 cache, 10 for the L2 cache,
 and 4 in L1 (using AVX).
}
\end{figure}

In case of the DAXPY example on the Intel SNB core,
these rules lead to the timeline diagram shown in
Fig.~\ref{fig:daxpy_serial}.
In this particular case the overall execution time is dominated by
data transfers, even if the data comes from the L1 cache. 

In order to be able to discuss the ECM model more efficiently we
introduce a shorthand notation that summarizes the most relevant
information from timeline diagrams such as Fig.~\ref{fig:daxpy_serial}.
We write the predicted cycle counts for $T_\mathrm{OL}$ and
$T_\mathrm{nOL}$, separated by ``$\olsep$''. After $T_\mathrm{nOL}$ we
attach the individual contributions for the different memory hierarchy levels
(which make up $T_\mathrm{data}$), separated by ``$\nolsep$'':
\bq
\ecm{T_\mathrm{OL}}{T_\mathrm{nOL}}{T_\mathrm{L1L2}}{T_\mathrm{L2L3}}{T_\mathrm{L3Mem}}{}
\eq
For example, the \ecmm\ for DAXPY would be written
as \ecm{4}{4}{6}{6}{13}{\cycles}.
Cycle predictions for data sets fitting into any particular cache level can then be
easily calculated from this representation by adding up the
appropriate contributions from $T_\mathrm{data}$ and $T_\mathrm{nOL}$,
and finding the maximum of that number and $T_\mathrm{OL}$ according
to (\ref{eq:T}). For instance, the prediction for data in L2 will be
$\max\left(4,4+6\right)\,\cycles=10\,\cycles$. As a shorthand notation
for these predictions we use a similar format but with ``$\epsep$''
as the delimiter. For DAXPY
this would read as
$T_\mathrm{ECM}=\ecmp{4}{10}{16}{29}{\cycles}$.

As a further example that does not exhibit the strong data transfer
dominance of DAXPY we pick a double precision vector summation:
\begin{lstlisting}
 for(i=0; i<N; ++i)
  s = s + a[i];
\end{lstlisting}
We construct the \ecmm\ for this kernel in four cases: naive, scalar,
SSE, and AVX vectorization on a SNB core. In all but
the ``naive'' case we assume that
appropriate modulo unrolling has been applied (at least 3-way due to the
3-cycle ADD latency) on top of SIMD vectorization to allow for filling the ADD
pipeline despite the register dependency. The ``naive'' code is scalar
and does not use modulo unrolling, so the full latency must be paid for every
ADD instruction. The results are shown in Table~\ref{tab:sumpred}.
\begin{table}[tb]
\renewcommand{\arraystretch}{1.3}
\centerline{\begin{tabular}{rccc}\hline
 case   & \ecmm\ [\cycles] & prediction [\cycles] & measurement [\cycles]\\\hline
 naive  & \ecm{24}{4}{2}{2}{4.3}{} & \ecmp{24}{24}{24}{24}{} & \ecme{24}{24}{24}{27}{}\\
 scalar & \ecm{8}{4}{2}{2}{4.3}{}  & \ecmp{8}{8}{8}{12}{}  & \ecme{8.1}{8.9}{11}{18}{}\\
 SSE    & \ecm{4}{2}{2}{2}{4.3}{}  & \ecmp{4}{4}{6}{10}{}  & \ecme{4.1}{5}{7.2}{15}{}\\
 AVX    & \ecm{2}{2}{2}{2}{4.3}{}  & \ecmp{2}{4}{6}{10}{}  & \ecme{2.1}{4.9}{7.2}{14}{}\\\hline
\end{tabular}\smallskip}
\caption{\label{tab:sumpred}\ecmm\ and cycle predictions on a SNB
  core for the vector summation loop in double precision.}
\end{table}
One can immediately see that the naive code is executing so slowly that 
it does not ``feel'' the memory hierarchy even if the data is in 
memory. The scalar code is dominated by core
execution down to the L3 cache and starts to degrade in performance
only when the data comes from memory. Furthermore, although SSE
vectorization yields a significant speedup in L1 and L2 cache, the gain
is only about 20\% in memory. Finally, going from SSE to AVX pays
off only if the data is in the L1 cache. 
These predictions are in line with the
general observation that the performance of ``slow,'' i.e.,
unoptimized code is largely independent of the position of data in the
memory hierarchy.

\subsubsection{Conclusions from the \ecmm}

The \ecmm\ explains accurately why the theoretical bandwidth limits of
all memory hierarchy levels except the L1 cache cannot be hit with a
single thread. Even with a highly efficient, bandwidth-demanding loop
kernel the non-overlapping contributions to $T_\mathrm{data}$ from the
higher memory levels prevent full bandwidth utilization. If multiple
threads are used, bandwidth saturation is possible.
See Sect.~\ref{sec:scaling} for scalability predictions.

Once the execution time for the basic work unit is known from the
\ecmm, one can calculate the single-core performance by dividing work
by time: $P_\mathrm{ECM}=W/T$. Any appropriate unit of ``work'' will
do, be it \flops, iterations, ``lattice site updates," etc.
In case of the scalar summation, the performance in \FS\ on
a SNB core
is
\bq
P = \frac{8\,\flops\cdot f}{\ecmp{8}{8}{8}{8+4.3f/f_0}{\cycles}}\eos
\eq
Here we allow for a modification of the CPU clock speed
away from the base frequency $f_0$.
At the nominal clock speed of $f_0=2.7\,\GHZ$ we get
$P(f_0)=\ecmp{2.7}{2.7}{2.7}{1.8}{\GFS}$. The model shows 
clearly that the single-core performance changes with the clock frequency
even if the data is in memory: $P(1.6\,\GHZ)=\ecmp{1.6}{1.6}{1.6}{1.2}{\GFS}$.
The larger the relative share of $T_\mathrm{L3Mem}$ versus the other
contributions, the weaker this dependence will be.

\subsubsection{Chip-level scaling and comparison with\\Roof\/line}\label{sec:scaling}

For describing the chip-level performance scaling we assume that the
single-core performance scales linearly with the number of cores until a bottleneck is
hit~\cite{Suleman:2008}. In case of the modern Intel processors the
only bottleneck is the memory bandwidth, which means that an upper
performance limit is given by the Roof\/line prediction for memory-bound
execution: $P_\mathrm{BW} = b_\mathrm S/B_\mathrm{C}$, where $B_\mathrm C$ is the
code balance (inverse computational intensity) 
of the loop code and $b_\mathrm{S}$ is the
memory bandwidth (one may certainly consider other bottlenecks, such
as a non-scalable L3 cache like on the Intel Westmere processors).
The performance scaling for $n$ cores is thus described by
\bq\label{eq:ecmsat}
P(n)=\min\left(nP_\mathrm{ECM}^\mathrm{mem},\frac{b_\mathrm S}{B_\mathrm C}\right)
\eq
if $P_\mathrm{ECM}^\mathrm{mem}$ is the \ecmm\ prediction if 
the data is in main memory. 

This expression allows us to draw a comparison between the \ecmm\ and the \rlm.
The \rlm\ considers only a single data transfer bottleneck. This could
be main memory access but also any other memory hierarchy level.
Constructing the \rlm\ for any core count thus requires measured maximum
bandwidth numbers for all memory levels and all core counts as input
parameters, since none of these values can be predicted by the model. 
Due to the overlapping assumption of the \rlm, the predicted 
single-core performance is then
\bq\label{eq:rl1}
P^{\mathrm{Roof}}(n) = \min_i\left(nP_\mathrm{max}^\mathrm{core},\frac{b_{\mathrm S,i}(n)}{B_{\mathrm{C},i}}\right)\eos
\eq
Here, $P_\mathrm{max}^\mathrm{core}$ is the maximum in-core sequential
performance, i.e., with all data coming from the L1 cache (one may use 
the arithmetic peak performance here and treat the L1 cache as an additional
memory level).
It
corresponds to $T_\mathrm{core}$ as used in the \ecmm. $b_{\mathrm S,i}(n)$ 
and $B_{\mathrm{C},i}$ are the measured maximum bandwidth
and code balance, respectively, at $n$ cores and in memory level
$i$ (excluding L1). There are two limiting cases where the \rlm\ and \ecmm\ predictions
coincide on the processor architectures considered here: 
\begin{itemize}
\item $T_\mathrm{core}$ is large compared to all contributions from
  data transfers. Then, if $n$ is large enough, the memory bottleneck
  may become relevant, but this is described in the same way by both
  models. One typical example would be a kernel that is absolutely
  dominated by long-latency operations such as divides or square
  roots, like the ``divide triad'' kernel analyzed
  in~\cite{hager:cpe}.
\item The loop code has single-core data transfer and execution
  characteristics that make its Roof\/line-based performance estimate
  coincide with the measured streaming benchmark performance used for
  obtaining the $b_{\mathrm S,i}(n)$. We will show an example for this
  in Sect.~\ref{sec:j2d_blocking}.
\end{itemize}
If, however, in-core execution time and data transfer times are
comparable but much larger than those of the streaming benchmark,
the \rlm\ is not able to describe the performance
characteristics accurately. See Sect.~\ref{sec:longstencil} for
an example.

 
According to (\ref{eq:ecmsat}), the performance will saturate at $n_\mathrm{S}$ cores:
\bq
n_\mathrm S = \left\lceil\frac{b_\mathrm S/B_\mathrm C}{P_\mathrm{ECM}^\mathrm{mem}}
	\right\rceil 
 = \left\lceil\frac{T_\mathrm{ECM}^\mathrm{mem}}{T_\mathrm{L3Mem}}  
	\right\rceil \eos
\eq
The model confirms the well-known effect that ``slow'' loop code needs
more cores to reach saturation. For example, the AVX summation code
has a predicted single-core performance of
$P_\mathrm{ECM}^\mathrm{mem}=2.1\,\GFS$ at nominal clock speed and it
will saturate at three cores, while the naive, non-pipelined code with
$P_\mathrm{ECM}^\mathrm{mem}=0.9\,\GFS$ will require six. Hence, in
this case a low code quality can be ``healed'' by using more
resources.\footnote{If time to solution is the only relevant metric,
 this attitude may be justified. However, it is not acceptable in
 view of energy consumption. See~\cite{hager:cpe}, and references
 therein.} This would not be possible at a clock speed of 1.6\,\GHZ,
since the slow code would then need ten cores to saturate, but the CPU
only has eight.

\section{Case study: 5-point stencil}
\label{sec:j2d}

\begin{table}[tb]
\centerline{\renewcommand{\arraystretch}{1.3}
\begin{tabular}{rccccc}\hline
LC &  \ecmm\ [\cycles]  & prediction [\cycles] &  \multicolumn{1}{m{1cm}}{\centering $P^\mathrm{mem}_\mathrm{ECM}$ [\MLUPS]} & $N_i<$ & $n_\mathrm S$\\\hline 
L1             &  \ecm{6}{8}{6}{6}{13}{}   & \ecmp{8}{14}{20}{33}{} & 659 & 683    & 3 \\
L2             &  \ecm{6}{8}{10}{6}{13}{}  & \ecmp{8}{18}{24}{37}{} & 587 & 5461   & 3 \\
L3             &  \ecm{6}{8}{10}{10}{13}{} & \ecmp{8}{18}{28}{41}{} & 529 & 436900 & 4 \\
---            &  \ecm{6}{8}{10}{10}{22}{} & \ecmp{8}{18}{28}{50}{} & 438 & N/A    & 3 \\\hline  
\end{tabular}\smallskip}
\caption{\label{tab:j2d_ecm}\ecmm, main memory code balance, predicted
  performance, required leading dimension, and saturation point for the 2D Jacobi kernel
  when the layer condition (LC) is fulfilled in different memory hierarchy
  levels, from L1 down to main memory on one SNB core.}
\end{table}
We use the arguably most simple non-trivial stencil variant 
to employ the \ecmm\ in a setting that is more complex than
the pure streaming kernels shown earlier. All concepts, however,
are generalizable to other stencils. One sweep (complete
update of all points) of the 2D five-point
Jacobi stencil looks as follows (double precision assumed):
\begin{lstlisting}
 for(j=1; j<%$N_j$%-1; ++j)
   for(i=1; i<%$N_i$%-1; ++i)
     b[j][i] = ( a[j][i-1] + a[j][i+1] 
               + a[j-1][i] + a[j+1][i] ) * s;
\end{lstlisting}
Unless otherwise noted we assume that the array dimensions are large
so that the arrays \verb.a. and \verb.b. do not fit in any cache. We
adopt ``lattice site updates per second'' (\LUPS), i.e., scalar inner
kernel iterations, as an appropriate performance metric. 

\subsection{Basic analysis}\label{sec:j2d_basic}

With AVX vectorization, one unit of work (eight \LUP{}s) comprises
eight loads, two stores, six adds, and two multiply instructions
(note that there will be no benefit from inner loop unrolling to
enable re-use of data along the leading dimension). Hence, 
$T_\mathrm{nOL}=8\,\cycles$ and $T_\mathrm{OL}=6\,\cycles$. The
code is thus limited by the LOAD pipeline throughput at the core
level.

The data transfer properties of the loop nest depend on the problem
size in relation to the available cache sizes. The target array
\verb.b. causes a traffic of two CLs across the complete
memory hierarchy due to the write-allocate transfer on every store
miss. The right hand side element \verb.a[j][i-1]. can always be
obtained from the L1 cache since it was loaded two inner iterations
before as \verb.a[j][i+1]., and \verb.a[j+1][i]. must be loaded from
memory since it was not used before within the sweep. Whether
\verb.a[j-1][i]. and \verb.a[j][i+1]. cause cache misses in a certain
cache level depends on whether three successive rows of the array
\verb.a. fit into the cache~\cite{datta09}. We call this the
\emph{layer condition} (LC). If the size of cache $k$ is $C_k$, 
it can be formulated as
\bq\label{eq:2dlayer}
3\cdot N_i\cdot 8\,\bytes < \frac{C_k}{2}\eos
\eq
The factor of $1/2$ is a rule of thumb: we assume that half the cache
size is available for storing layers. The number of layers is connected to
the \emph{stencil radius} in the outer dimension. If $r$ is the radius,
$(2r+1)$ layers have to fit.

If the layer condition is
fulfilled for cache level $k$, only \verb.a[j+1][i]. causes a cache
miss in this level, and is subsequently used three times from the
cache. The code balance is then $24\,\bytes/\LUP$ as seen from the
next memory hierarchy level (cache $k+1$, or main memory), because
only three data streams must be sustained. If the condition is
violated, only \verb.a[j][i-1]. comes from the cache and all other
elements cause misses, leading to five data streams and a code balance
of $40\,\bytes/\LUP$. Columns 2--4 in Table~\ref{tab:j2d_ecm} show a
summary of the \ecmm\ timing and performance predictions 
on a SNB core. In column 5 we give the
leading dimension below which the respective layer condition is
satisfied. This number was obtained by solving (\ref{eq:2dlayer}) 
for $N_i$ and using the
cache sizes of the SNB processor in Table~\ref{tab:arch}.

\subsection{Single-core performance vs.\ problem size}\label{sec:j2d_sc}

\begin{figure}[tb]
\centerline{\includegraphics*[width=0.9\linewidth]{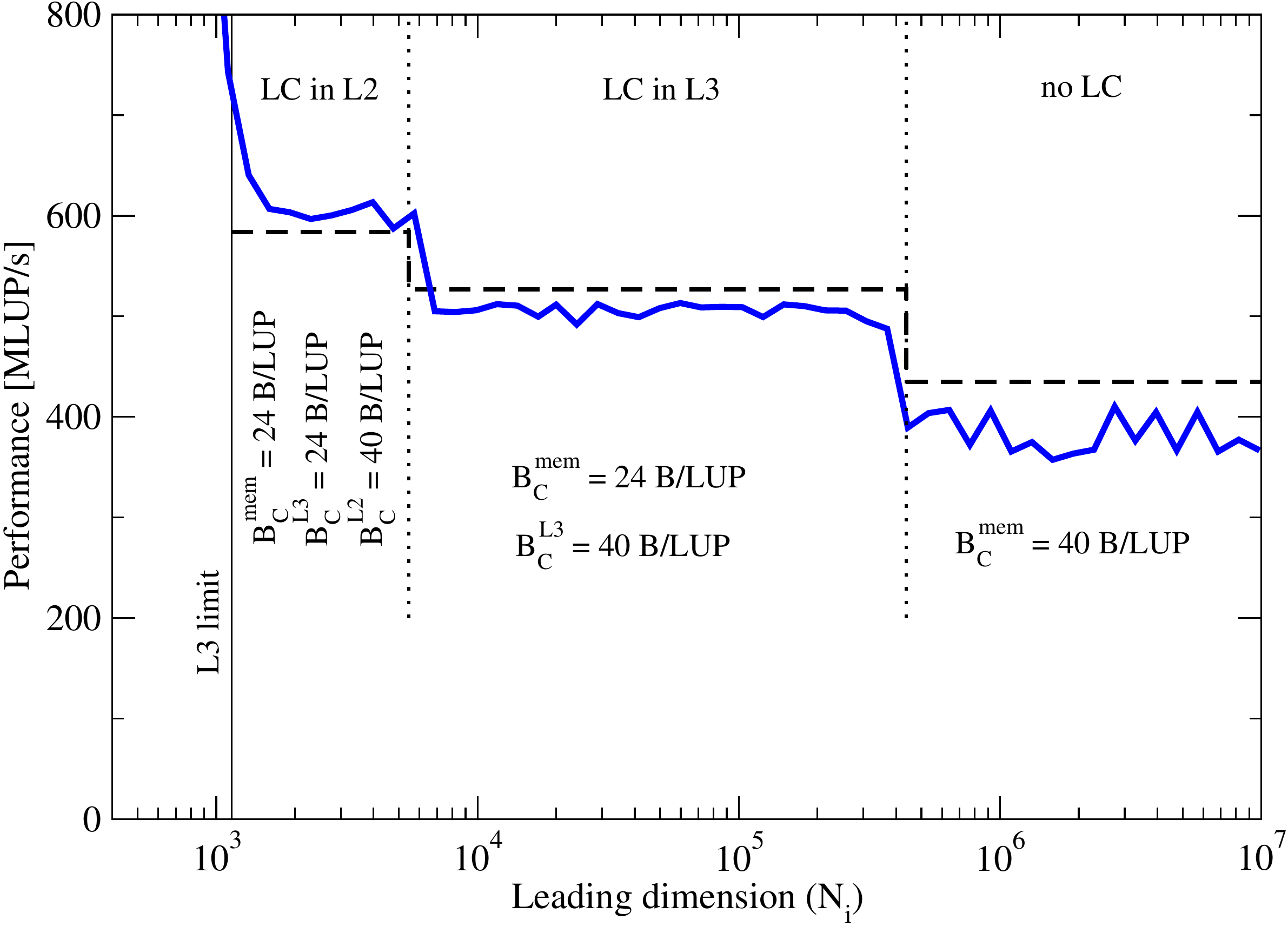}}
\caption{\label{fig:j2d_scan_phinally}In-memory performance vs.\ leading array
 dimension of the 2D Jacobi sweep on a SNB core. Up to a
  leading dimension of $2\times 10^4$ the domain is quadratic. Beyond 
  that, the overall problem size is kept constant. The dashed line is 
  the prediction from the \ecmm.}
\end{figure}

Figure~\ref{fig:j2d_scan_phinally}
shows the measured performance and the \ecmm\ prediction for 
a large range of memory-bound problem sizes. 
After the problem drops out to main memory at $N_i\approx 1145$
(quadratic domain), three phases
can be distinguished. These can be attributed to the layer condition
being satisfied in the L2 cache, the L3 cache, and not at all, 
corresponding to rows 2--4 in Table~\ref{tab:j2d_ecm}. An
L1 layer condition would require a leading dimension of less than
$683$, which leads to an in-cache problem size if $N_i=N_j$. For each
phase we also show the predicted code balance for different
levels of the memory hierarchy. These results indicate that the
\rlm\ may not be a reliable model for the single core, because the
performance changes considerably when the leading dimension grows
beyond $N_i=5461$ although $B_\mathrm{C}^\mathrm{mem}$ stays the same.
See below for more discussion on this issue.

The dashed line in Fig.~\ref{fig:j2d_scan_phinally} shows the
\ecmm\ prediction, which is in good agreement with the measurement in
all three phases. The model fits well within a 10\% margin, and it is
even slightly below the measurement if the layer condition is
fulfilled in L2 cache. We have observed this effect in many
situations where the pressure on the memory hierarchy gets smaller
when moving towards main memory (i.e., when the number of streams goes
down). One may speculate that the non-overlapping assumption of the
model is inaccurate in these cases; the deviation is generally small,
however.

\begin{figure}[tb]
\centerline{\includegraphics*[width=0.9\linewidth]{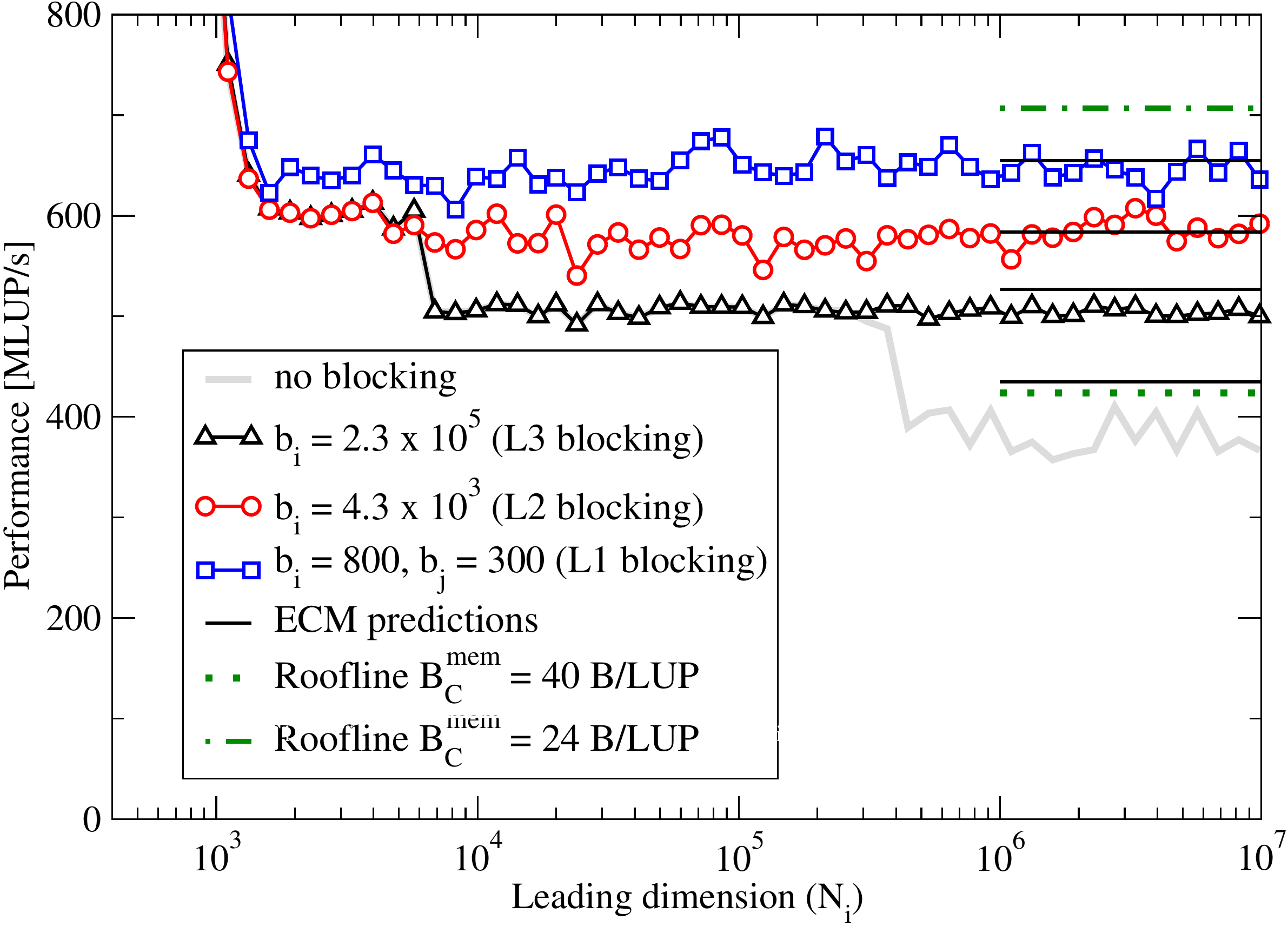}}
\caption{\label{fig:j2d_scan_phinally_all}Performance vs.\ problem size
  for the 2D Jacobi stencil
  as in Fig.~\ref{fig:j2d_scan_phinally} but with spatial blocking
  of the leading dimension with block size $b_i$. The four \ecmm\ predictions
  correspond to the rows of Table~\ref{tab:j2d_ecm}. In case of the L1 layer
  condition (squares) blocking in $j$ direction is required (see also
  Fig.~\ref{fig:j2d_scan_phinally_L1L}). The single-core Roof\/line predictions
  for the two in-memory code balance values are shown for comparison.
  Roof\/line does not yield different predictions across different
  blocking strategies.}
\end{figure}
Despite the performance fluctuations (especially in the phase with
$B_\mathrm C=40\,\bytes/\LUP$), the actual memory code balance
obtained by measuring the data volume with hardware performance events
is within 5\% of the predicted value.

\subsection{Spatial blocking optimization}\label{sec:j2d_blocking}

Spatial blocking (also called ``loop tiling'') is a well-known
optimization for loop nests that aims at improving temporal data
access locality. In the context of stencil solvers it amounts to
rearranging the order of the stencil updates so that some layer
condition is met, leading to a decrease in code balance. We can
predict the expected gain of spatial blocking on the single core by
setting up an \ecmm\ for different blocking strategies. This is
especially simple for the 2D Jacobi stencil as the layer condition
(\ref{eq:2dlayer}) only depends on the leading dimension. One must
conclude that it is sufficient to block the inner loop:
\begin{lstlisting}
 // inner block size = b_i
 for(is=1; is<%$N_i$%-1; is += %$b_i$%)
   for(j=1; j<%$N_j$%-1; ++j)
     for(i=is; i<min(%$N_i$%-1,is+%$b_i$%-1; ++i)
       b[j][i] = ( a[j][i-1] + a[j][i+1] 
                 + a[j-1][i] + a[j+1][i] ) * s;
\end{lstlisting}
The layer condition (\ref{eq:2dlayer}) now takes a modified form:
\bq\label{eq:2dlayer_b}
3\cdot b_i\cdot 8\,\bytes < \frac{C_k}{2}\eos
\eq
Hence, the block size $b_i$ determines whether the layer condition
will be satisfied in cache level $k$, and column 5 in
Table~\ref{tab:j2d_ecm} again provides the relevant thresholds.
Choosing $b_i$ appropriately is called ``blocking for cache level $k$.'' 
Figure~\ref{fig:j2d_scan_phinally_all} shows performance vs.\ leading
dimension for three block sizes and no blocking. These four cases
(squares, circles, triangles, gray line) correspond to the four
possible layer conditions in Table~\ref{tab:j2d_ecm}, and the \ecmm{}s
given there apply. The solid lines denote the ECM performance predictions. 
The dotted and dashed-dotted lines mark the \rlm\ predictions
derived from measured single-thread cache and memory
bandwidths.\footnote{STREAM COPY kernel: $b_\mathrm S^\mathrm{mem}=17\,\GBS$, $b_\mathrm S^\mathrm{L3}=34\,\GBS$, $b_\mathrm S^\mathrm{L2}=56\,\GBS$. The assumed
    in-core performance was the same as for the \ecmm.}
Since the \rlm\ predicts a memory-bound situation in all cases (independent
of the blocking), there are only two performance levels: a lower level
for $B_\mathrm C^{\mathrm{mem}}=40\,\bytes/\LUP$ and a higher level for
$B_\mathrm C^{\mathrm{mem}}=24\,\bytes/\LUP$. 
Although the \rlm\ cannot distinguish between different blocking strategies,
the best and worst case scenarios (L1 blocking and no blocking) are
covered with useful accuracy. Note, however, that this is a coincidence:
As the \ecmm\ shows, the runtime contributions are rather evenly distributed
across the memory hierarchy. The \rlm\ works in this case because
the benchmark used for taking the single-thread memory bandwidth limit has similar
characteristics (in fact, the \ecmm\ prediction for the STREAM COPY loop
is \ecmp{4}{10}{16}{29}{\cycles}, and the number of data streams is the same as
for the Jacobi 2D kernel with L1 blocking). 

The data shows that the \ecmm\ is able to describe the performance
impact of spatial blocking quite accurately, and that the 
gain from moving a layer condition up in the memory hierarchy (i.e.,
satisfying it in a higher cache level) can be leveraged also for larger
leading dimensions. In case of pure L1
blocking, however, the observed performance was at first lower than
expected.
A measurement of the memory data traffic revealed that 
the actual code balance was much higher than $24\,\bytes/\LUP$, 
which was caused by the hardware prefetcher:
If the inner block size $b_i$ is small, the leading array 
dimension is only traversed partially, but the hardware
prefetcher cannot foresee when the loop ends. 
It fetches cache lines from beyond the
actual end of the short inner loop block. Those cannot be used 
before the next inner block is updated, they are thus evicted 
prematurely and cause excess memory traffic. 
Additional blocking in the outer dimension
was added to correct this problem:
\begin{figure}[tb]
\includegraphics*[width=0.475\linewidth]{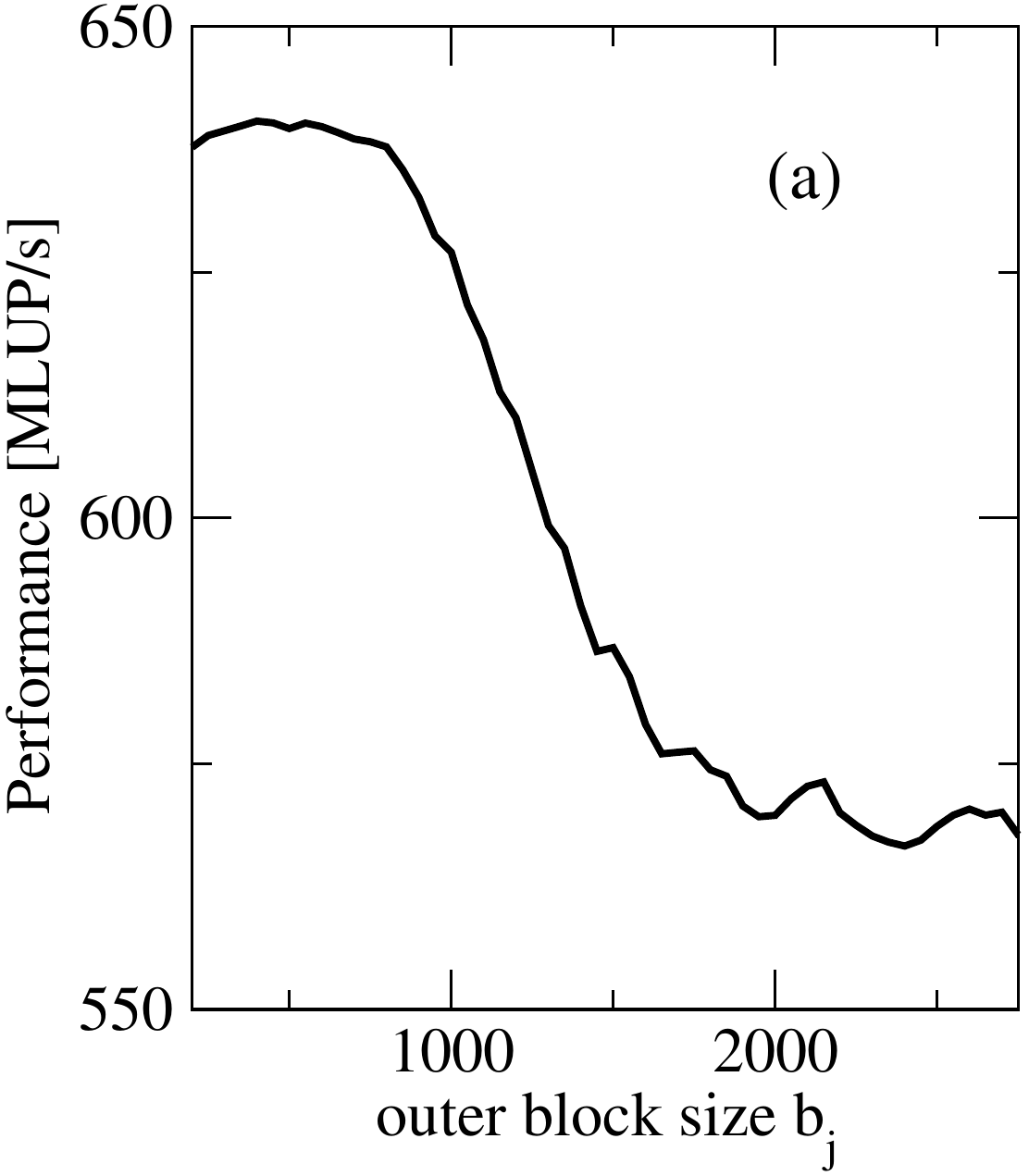}\hfill
\includegraphics*[width=0.475\linewidth]{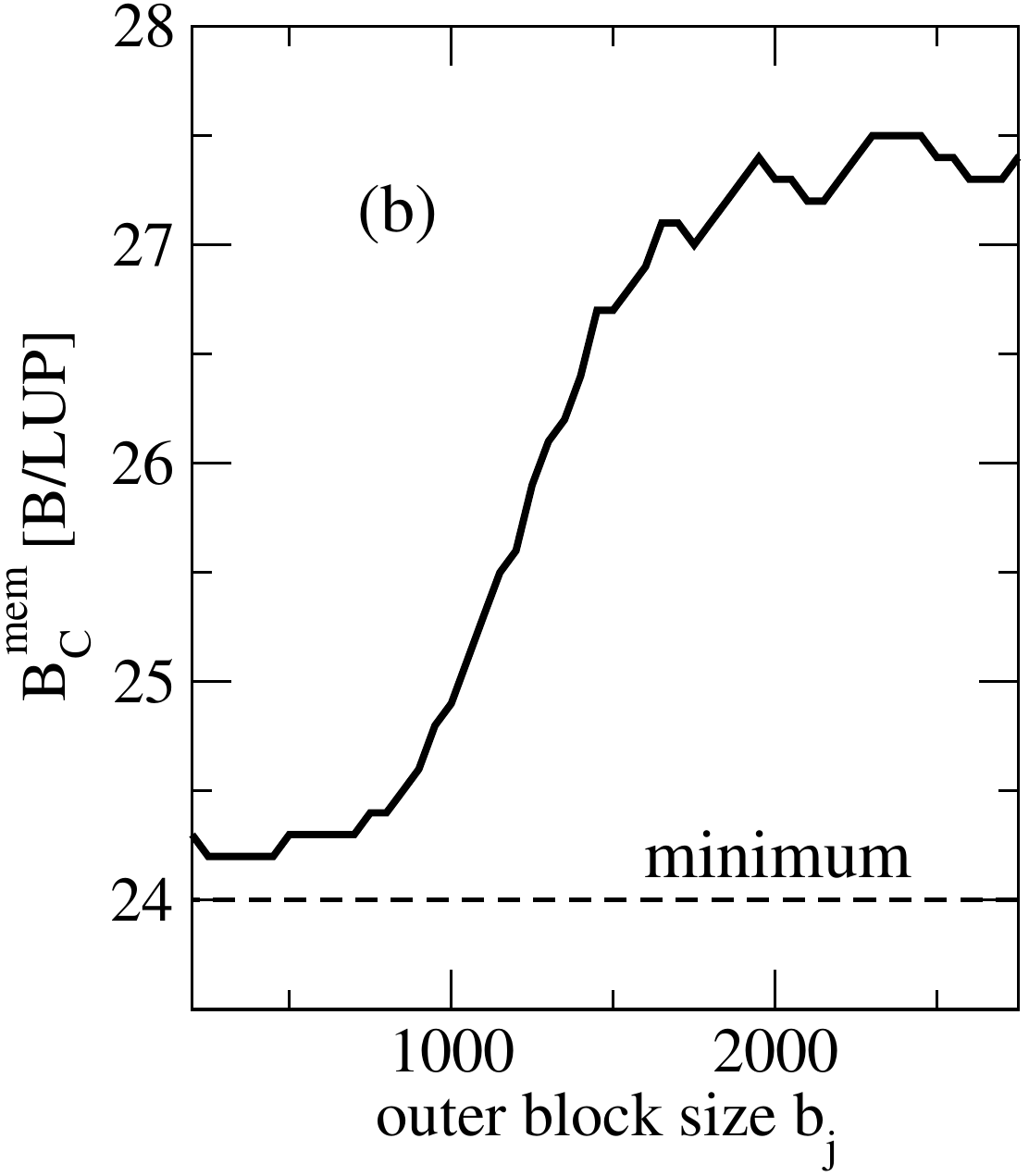}
\caption{\label{fig:j2d_scan_phinally_L1L}Influence of outer loop
  blocking on the 2D Jacobi stencil at a constant inner loop block
  size of $b_i=800$ and a constant problem size of $N_i=3.5\times
  10^4$ and $N_j=1.2\times 10^4$. (a) Performance vs.\ outer loop
  block size.  (b) Measured in-memory code balance vs.\ outer loop
  block size. }
\end{figure}
\begin{lstlisting}
 for(js=1; js<%$N_j$%-1; js += %$b_j$%)
   for(is=1; is<%$N_i$%-1; is += %$b_i$%)
     for(j=js; j<min(%$N_j$%-1,js+%$b_j$%-1); ++j)
       for(i=is; i<min(%$N_i$%-1,is+%$b_i$%-1); ++i)
         b[j][i] = ( a[j][i-1]+a[j][i+1] 
                   + a[j-1][i]+a[j+1][i] ) * s;
\end{lstlisting}
In Fig.~\ref{fig:j2d_scan_phinally_L1L} we show the effect of the
outer loop blocking with block size $b_j$ on the performance and
on the measured in-memory code balance at a constant inner block
size of $b_i=800$. It turns out that $b_j\approx 300$ alleviates
the prefetcher problem: blocks of size $b_i\times b_j$ are updated
consecutively along the leading dimension, and all excess cache lines
can be used. Note that one can determine the excess traffic, and hence
the prefetch distance, from the measured data volume at large 
$b_j$. We have calculated the prefetch distance to be
about 33 CL in our case, but this result is probably not
generic since the details of the prefetch mechanism are unknown.
Data overheads at block boundaries (of which eager prefetching is just
the most striking example) may be incorporated into the model
if the corresponding data volumes are known. 

\begin{figure*}[tb]
\includegraphics*[width=0.32\textwidth]{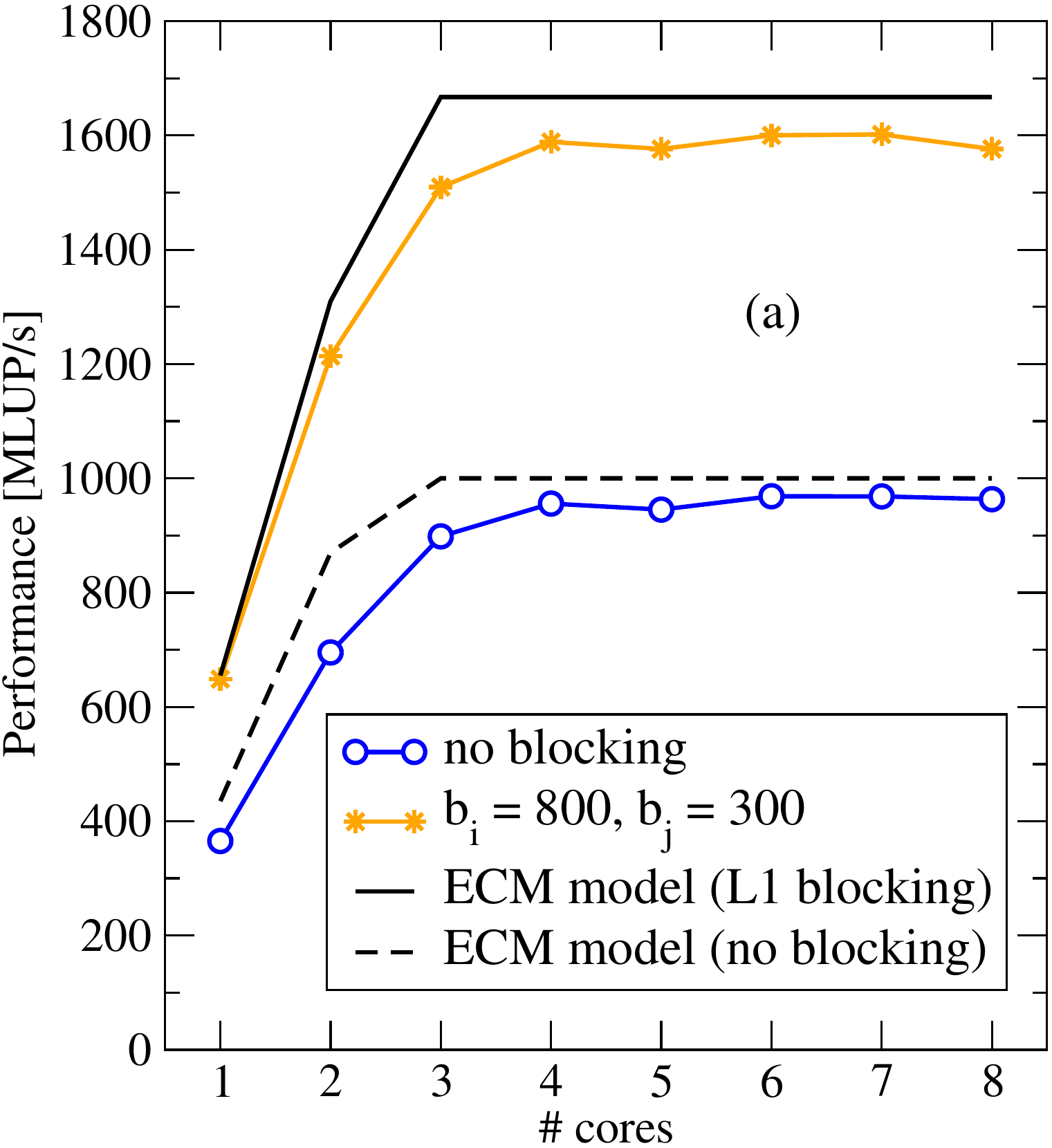}\hfill
\includegraphics*[width=0.32\textwidth]{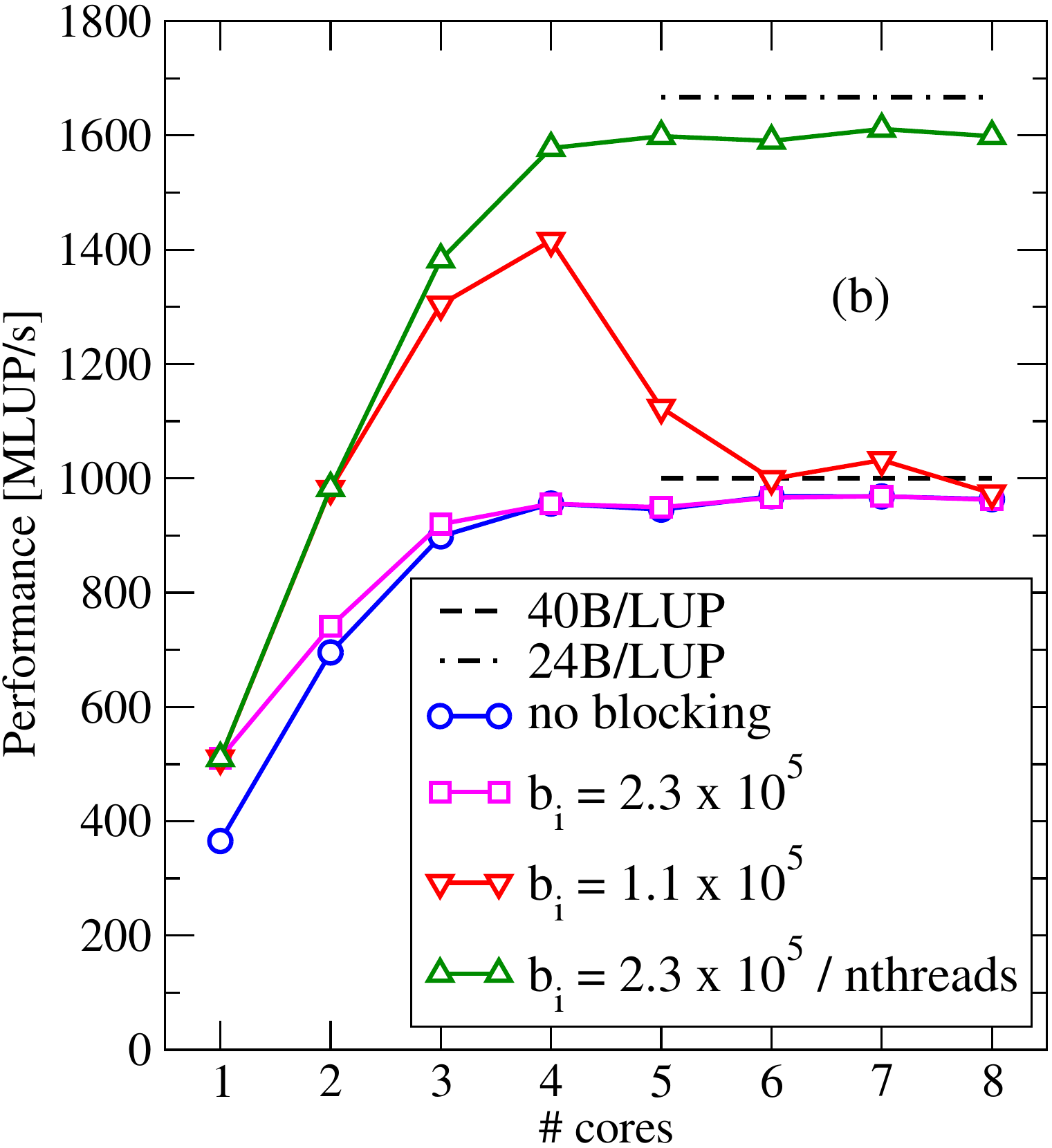}\hfill
\includegraphics*[width=0.32\textwidth]{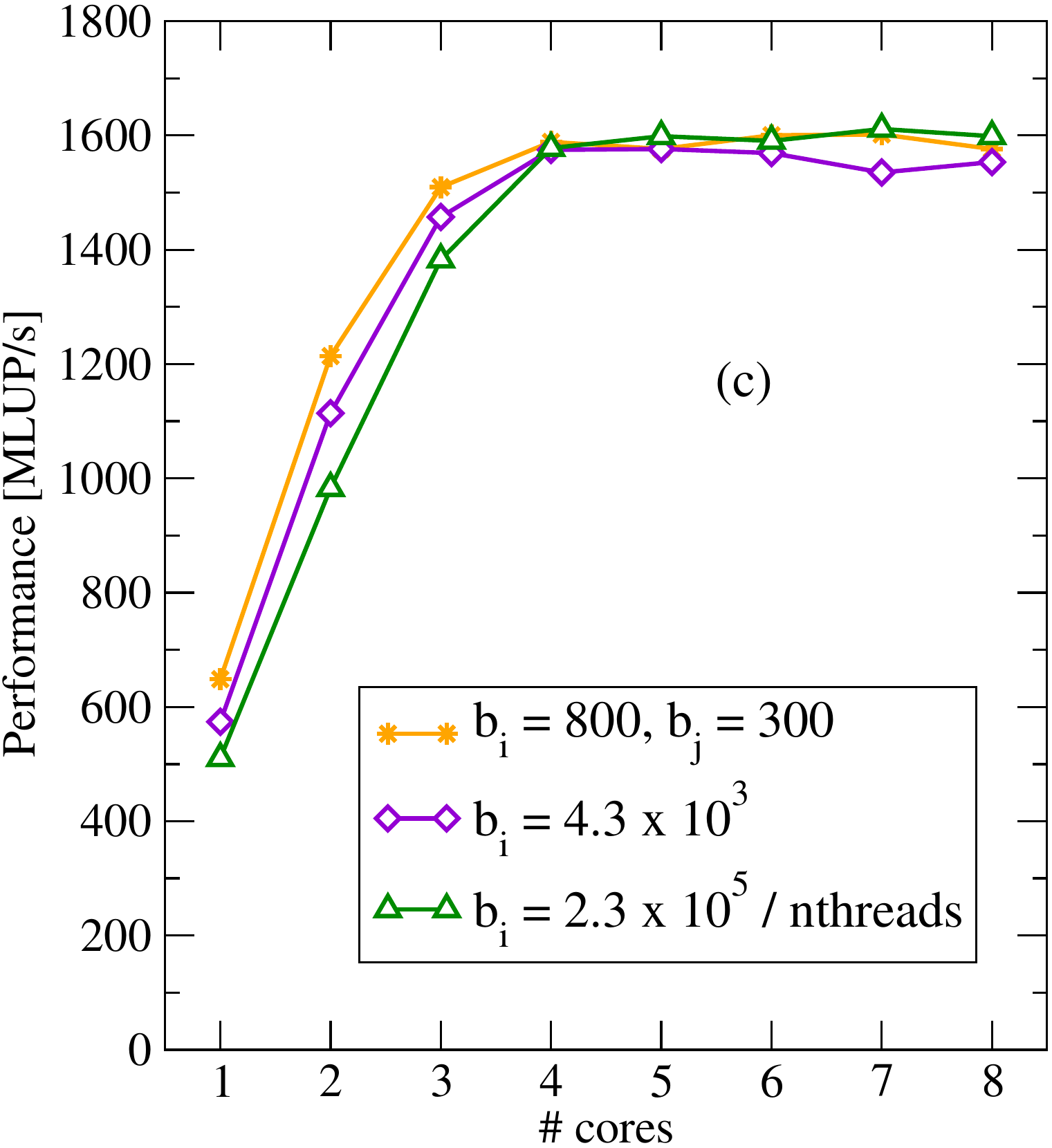}
\caption{\label{fig:j2d_scaling}Scaling of the Jacobi 2D kernel across the cores of a SNB 
  chip at $N_i=1.2\times 10^6$. (a) Comparison of \ecmm\ and measurement for 
  L1 blocking and no blocking. (b) Choosing the right block size $b_i$ for the L3 layer 
  condition. (c) Comparison of saturation behavior for L1, L2, and L3 blocking.}
\end{figure*}
Although all
considered problem sizes are bigger than the L3 cache, the execution
is far from memory bound. Even when blocking for the L1 cache (squares
in Fig.~\ref{fig:j2d_scan_phinally_all}), the memory bandwidth as
calculated from the product of performance and code balance is barely
$17\,\GBS$ out of a maximum of over $40\,\GBS$.  

Since the code balance is already at its possible minimum of
$24\,\bytes/\LUP$ across all cache levels, spatial blocking cannot
improve the single-core performance any further. The \ecmm\ prediction
in the first row of Table~\ref{tab:j2d_ecm} shows that if one
were able to reduce the core time by a factor of two (from 8 to 4
cycles) by some advanced optimization technique such as register
blocking, single-core performance would improve by a factor of
$33/(33-4)=1.14$.

\subsection{Multi-core scaling}

When going from single-threaded to multi-threaded execution on a
multicore chip, two architectural features impact the performance
behavior: (i) the memory bandwidth bottleneck and
(ii) the shared outer-level cache (if present). 
In Fig.~\ref{fig:j2d_scaling} we show performance scaling across the 
cores of a SNB socket at $N_i=1.2\times 10^6$.
Two-dimensional L1 blocking was only used where necessary.
In all
cases (blocked or not), the $j$ loop was parallelized with OpenMP
and static loop scheduling. This ensures that the prefetcher problem
described in the previous section remains solved, since each thread
works on a consecutive string of blocks in the leading dimension. 

Figure~\ref{fig:j2d_scaling}a compares the \ecmm\ and scaling predictions
with measurements for the two extreme cases of no blocking (circles)
and L1 blocking (stars). The general scaling behavior is described 
well, although there is a deviation in the vicinity of the saturation 
point, which we attribute to changes in the prefetching strategy
when the memory interface is almost fully utilized~\cite{intelopt}:
For instance, cache lines are only prefetched into L3 and the 
prefetch distance is smaller. This leads to an additional latency 
penalty, which is not part of the \ecmm\ (but could be included).
The saturation level is slightly smaller than predicted by the code 
balance, but the deviation is well below 10\%.

In Fig.~\ref{fig:j2d_scaling}b we study the influence of the block
size on the scalability when blocking for the L3 cache. At
$b_i=2.3\times 10^5$ (squares) the single-core speedup versus the
non-blocked version (circles) is as expected, but performance
saturates at a level that indicates a code balance of
$40\,\bytes/\LUP$. Reducing the block size to $b_i=1.1\times 10^5$
shows no improvement on the single core but leads to much higher
performance up to four cores. Beyond four cores, however, performance
degrades and settles again at the same level as before. The reason
for this behavior is the shared L3 cache, which requires a modified
layer condition:
\bq\label{eq:2dlayer_mt}
3\cdot b_i\cdot n\cdot 8\,\bytes < \frac{C_3}{2}\eos
\eq
The $n$-dependent block size ensures that three layers \emph{per
 thread} fit into the cache, and leads to the saturation performance
expected from a code balance of $24\,\bytes/\LUP$ (triangles). Note
that block size adaptivity is not required when blocking for
core-private caches, since their aggregate size scales with the number
of cores.

Finally we need to comment on the influence of the chosen blocking
strategy (L1, L2, or L3) on the scalability. Technically the
\ecmm\ predicts different saturation points for L3 blocking (4 cores)
and L1 or L2 blocking (3 cores), as shown in the last column of
Table~\ref{tab:j2d_ecm}. In reality they display very similar
saturation behavior for the Jacobi 2D kernel (see
Fig.~\ref{fig:j2d_scaling}c), although the general trend to saturate
earlier with faster single-core code can certainly be observed. If
maximum performance is the only significant metric, it is irrelevant
in practice which blocking strategy is chosen as long as the in-memory
code balance is at its minimum of $24\,\bytes/\LUP$. Saturating
earlier, however, has some advantages: ``expendable'' cores can be put
into a low-power mode, saving energy (``concurrency throttling''), or
they can be used for other tasks such as asynchronous communication.

Note that the use of non-temporal stores (also called ``streaming
stores'') may reduce the in-memory code balance from $24\,\bytes/\LUP$
to $16\,\bytes/\LUP$ by ignoring the cache hierarchy on a store miss,
saving the write-allocate transfers.  However, their implementation on
Intel processors cannot be described by a simple throughput
assumption.  Therefore it is unclear as yet how to incorporate the
non-temporal store instructions in the \ecmm.

\section{Case study: uxx stencil}\label{sec:uxx}

The ``uxx'' stencil \cite{DBLP:journals/procedia/ChristenS12}
is a part of a simulation code for 
dynamic rupture and earthquake wave propagation. The basic
loop nest without any blocking optimizations looks as follows:
\begin{lstlisting}
#pragma omp parallel for schedule(static) \
                         private(d)  
for(int k=2; k<=N-1; k++){
 for (int j=2; j<=N-1; j++){
  for (int i=2; i<=N-1; i++){
   d = 0.25*(%d1%[ k ][j][i] + %d1%[ k ][j-1][i]
           + %d1%[k-1][j][i] + %d1%[k-1][j-1][i]);
   u1[k][j][i] = u1[k][j][i] + (dth%/%d)
   *( c1*(xx[ k ][ j ][ i ]-xx[ k ][ j ][i-1])
    + c2*(xx[ k ][ j ][i+1]-xx[ k ][ j ][i-2])
    + c1*(xy[ k ][ j ][ i ]-xy[ k ][j-1][ i ])
    + c2*(xy[ k ][j+1][ i ]-xy[ k ][j-2][ i ])
    + c1*(xz[ k ][ j ][ i ]-xz[k-1][ j ][ i ])
    + c2*(xz[k+1][ j ][ i ]-xz[k-2][ j ][ i ]));
}}}}
\end{lstlisting}
The code is used with either single or double precision, and it contains
a divide operation in the inner loop. Without any
guidance by performance modeling one would probably try to
eliminate the divide (which is non-trivial) because 
``divides are expensive.''
\begin{table}[tb]
\centerline{\renewcommand{\arraystretch}{1.3}%
\begin{tabular}{rcc}\hline
version   & \ecmm\ [\cycles] & prediction [\cycles] \\\hline
DP        & \ecm{84}{38}{20}{20}{26}{} & \ecmp{84}{84}{84}{104}{} \\
SP        & \ecm{45}{38}{20}{20}{26}{} & \ecmp{45}{58}{78}{104}{} \\
DP noDIV  & \ecm{41}{38}{20}{20}{26}{} & \ecmp{41}{58}{78}{104}{} \\\hline
\end{tabular}\smallskip}
\caption{\label{tab:uxx_ecm}\ecmm\ and cycle predictions for the uxx
  stencil code in double precision (DP) and single precision (SP),
  respectively, with L3 blocking. The unit of work is $8\,\LUP$s
  for DP and $16\,\LUP$s for SP. The ``noDIV'' version performs a
  multiply instead of a divide in the inner loop.}
\end{table}


\subsection{Analysis and \ecmm}

The loop code as generated by the compiler is quite complex, so we
employ the IACA tool. At double precision (DP) the throughput analysis 
reports the divider
to be the bottleneck on the core, causing a core time of
$T_\mathrm{core}=T_\mathrm{OL}=84\,\cycles$ per CL (eight
iterations) due to the very slow 42-cycle throughput for the divide
instruction (\verb.vdivpd.). The non-overlapping part is
$T_\mathrm{nOL}=38\,\cycles$, caused by the loads to the arrays
\verb.xx., \verb.xy., \verb.xz., \verb.d1., and \verb.u1.. With single
precision (SP) the machine code does not contain any divide at all:
the compiler uses the \verb.vrcpps. instruction instead, which
provides a low-precision ($11\,\bits$) but fully vectorized and
pipelined reciprocal, and employs a subsequent Newton-Raphson step and
a multiply to perform a 22-\bit\ divide~\cite{intelopt}. This results
again in $T_\mathrm{nOL}=38\,\cycles$ as the load instructions are the
same as with DP. The overlapping part is $T_\mathrm{OL}=35\,\cycles$,
caused by add, multiply, and 16-byte move instructions. The code is
close to the limit of four micro-ops per cycle on the SNB,
which is why IACA reports a front-end bottleneck and an overall
throughput of $T_\mathrm{core}=45\,\cycles$. Since front-end stalls
overlap with the transfer time, the seven extra cycles have no
significance and will be ignored in the following.

Calculating the transfer time requires an analysis of relevant layer
conditions. There are two types of layers in a 3D stencil
code: one-dimensional layers along the leading dimension (rows along
$i$ of size $N$) and layers that span the extent of the inner
two-level loop nest (size $N\times N$). At relevant problem
sizes only the latter will matter, and we can safely ignore the
``row conditions'' since they will be automatically fulfilled
in the L1 cache. The only data structures which are subject to
layer conditions are thus \verb.xz. and \verb.d1., because 
their stencil radius is larger than zero in the outer ($k$) 
dimension. The cache must hold as many layers as there are distinct 
outer index references, i.e., two for \verb.d1. (which is accessed
in layers $k$ and $k-1$) and four for 
\verb.xz. (which is accessed in layers $k-2$ to $k+1$). 
Hence, the layer condition for $n$ threads in the 
shared L3 cache and blocking in the $j$ dimension is
\bq\label{eq:uxx_layer}
(4+2)\cdot N\cdot b_j\cdot n\cdot \left\{
\begin{array}{c}
4\,\bytes~ \text{(SP)}\\
8\,\bytes~ \text{(DP)}
\end{array}
\right\} < \frac{C_3}{2}\eos
\eq
One may certainly consider higher cache levels or block the inner
dimension as well, but we omit this discussion here since it would
not provide new insights. 

With L3 blocking, the transfer time will be independent of the
precision.
Six consecutive data streams hit the memory interface per
thread: one each for \verb.xx., \verb.xy., \verb.xz., and \verb.d1.,
and two for \verb.u1.. The code balance in memory is thus
$B_\mathrm{C}^\mathrm{mem}=24\,\bytes/\LUP$ in SP and
$48\,\bytes/\LUP$ in DP. Assuming that the higher caches are too small
to hold the layers, the L3 cache will be hit by ten streams per
thread.

\begin{figure}[tb]
\includegraphics*[height=0.68\linewidth]{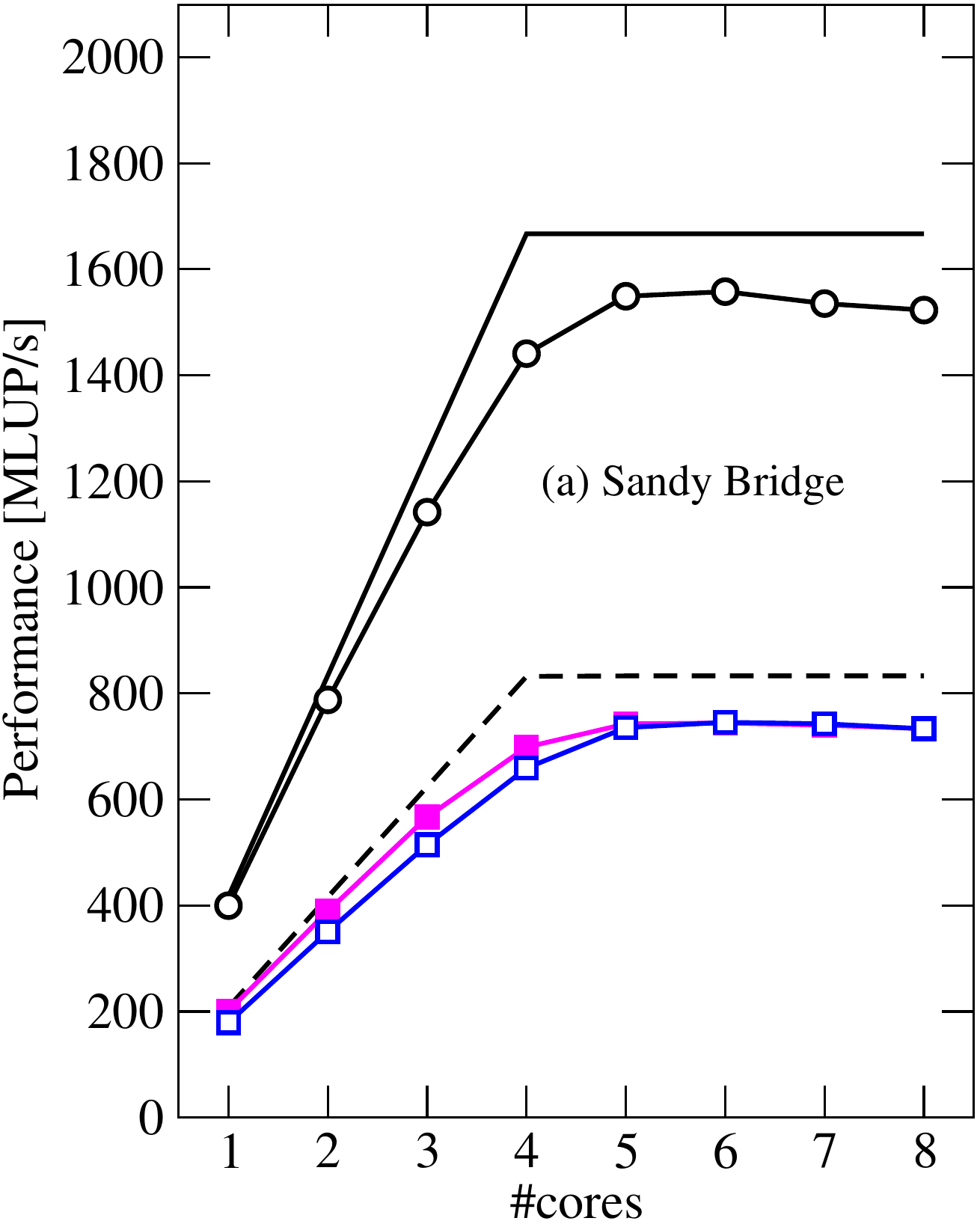}\hfill
\includegraphics*[height=0.68\linewidth]{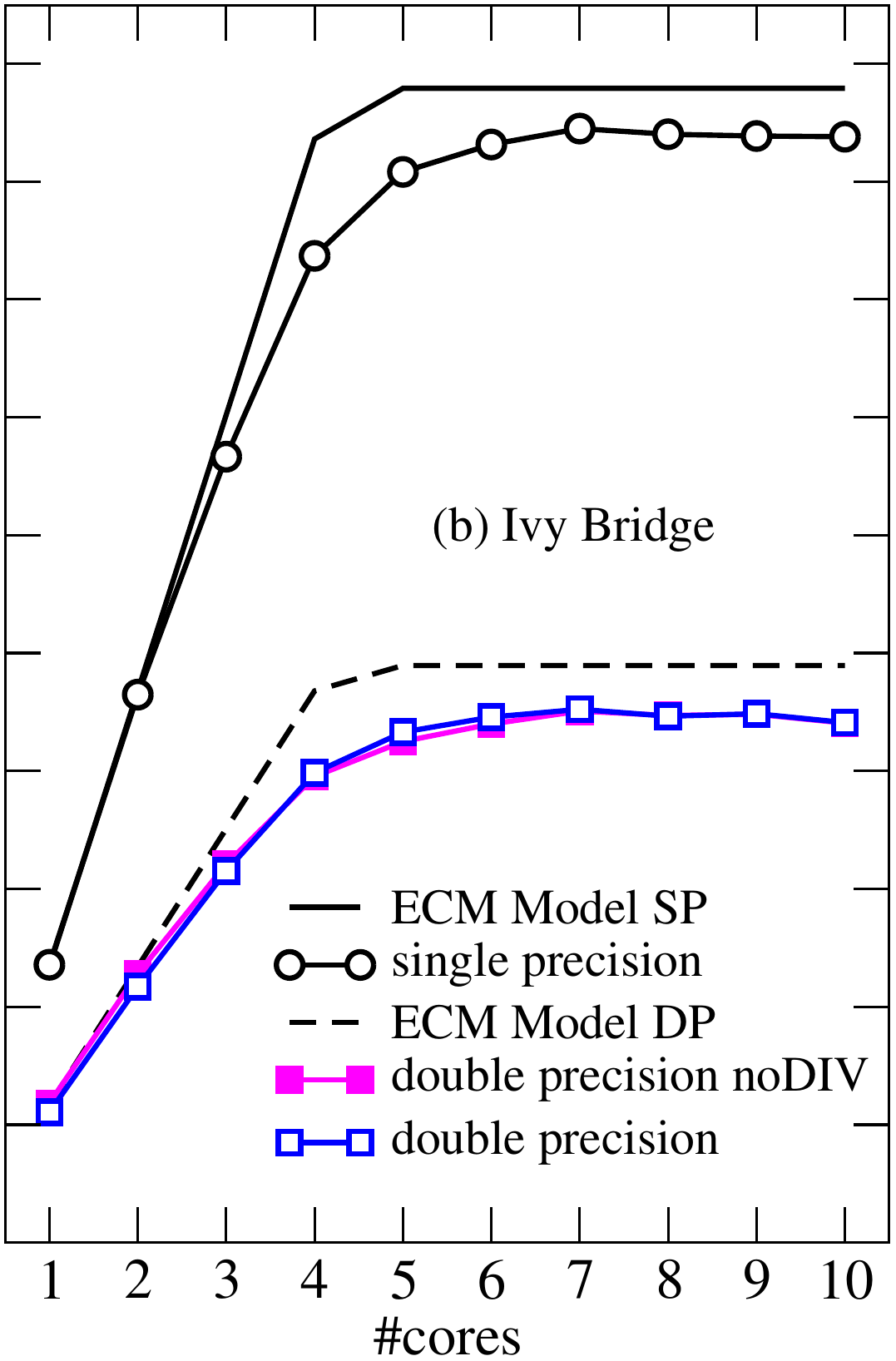}
\caption{\label{fig:uxx-scaling}Performance scaling and
  \ecmm\ predictions of the uxx stencil benchmark with L3 blocking
  at a problem size of
  $276^3$ grid points across the cores of one SNB socket (left)
  and one IVB socket (right)
  in single and double precision.}
\end{figure}

The analysis in the previous sections now enables us to construct 
the \ecmm\ for the uxx stencil. The first two rows in Table~\ref{tab:uxx_ecm}
show the model and the cycle predictions for SP and DP. For comparison
we also include a special DP version (``noDIV'') in which the divide
as substituted by a multiplication. 

The \ecmm\ yields the same cycle prediction per unit of work in main
memory for all versions ($104\,\cycles$), hence we expect a factor of
two in performance between SP and DP. Also, all versions should
saturate at four cores. Probably the most striking result is that the
presence or absence of the divide does not make a difference as
\bq\label{eq:uxx_dd}
T_\mathrm{data}+T_\mathrm{nOL}>T_\mathrm{OL}
\eq 
despite the
low-throughput divide. Going to great lengths in trying to eliminate
it from the code will thus not be rewarded with performance
improvements. These predictions are confirmed by the performance data
shown for one SNB socket in Fig.~\ref{fig:uxx-scaling}a. There
is actually a slight speedup for the ``noDIV'' version below
saturation. Using the latency analysis in the IACA tool revealed that
the critical code path through the loop body has a worst case
execution time of $192\,\cycles$. Usually the throughput prediction is
much closer to the measured runtime since the out of order logic can
fill a lot of pipeline bubbles across the independent loop iterations.
In this case, however, the critical path is so long that (\ref{eq:uxx_dd})
is just barely satisfied (in fact there is a 10\% deviation from
the model on the single core). Removing the divide shortens the
critical path so much that (\ref{eq:uxx_dd}) can be easily met,
and the \ecmm\ prediction is very precise. For comparison we present
the results on an Intel Ivy Bridge CPU ($3.0\,\GHZ$, $b_\mathrm S=47\,\GBS$)
in Fig.~\ref{fig:uxx-scaling}b. This architecture is quite similar
to Sandy Bridge apart from the higher memory
bandwidth and the faster divide throughput ($28\,\cycles$ instead of
$42\,\cycles$ for a double precision divide with AVX). The accuracy
of the model is even better here because of the shorter critical path
in double precision. The \ecmm\ in this case is \ecm{56}{38}{20}{20}{25}{\cycles},
leading to a prediction of \ecmp{56}{58}{78}{103}{\cycles}. The
divide instruction has no impact if the data does not fit into
the L1 cache.

\subsection{Optimization opportunities}

One typical stencil optimization that goes beyond spatial blocking is
\emph{temporal blocking}, which can dramatically reduce the in-memory
code balance by performing multiple updates on each grid point in
cache before it gets evicted. The \ecmm\ can provide upper limits to
the expected benefit of temporal blocking. In case of uxx, optimal
temporal blocking for the L3 cache would completely remove the memory
transfer time of $T_\mathrm{L3Mem}=26\,\cycles$, resulting in a 24\%
(DP) or 33\% (SP) speedup on the single core. The next bottleneck will
then be the divide for DP ($84\,\cycles$) and the L3 transfer time for SP
($78\,\cycles$), respectively, but both of these are purely
core-local. The true potential of temporal blocking is thus not 
in the speedup on the single core but in the removal of the memory
bandwidth bottleneck, allowing for scalable performance with an upper limit
of over $2000\,\MLUPS$ even for DP and still including the divide.
Even in this case the benefit from temporal blocking would by far
outweigh the speedup gained from removing the divide.  

See \cite{kronawitter14} for a case study of temporal blocking with
a 3D Jacobi smoother supported by an \ecmm\ analysis. 

\section{Case study: 3D long-range stencil}
\label{sec:longstencil}

A large stencil radius in the outer grid dimension leads to a high
pressure on the cache levels since they have to hold a large number of
layers. As an example we consider a constant coefficient long-range
stencil code in single precision with a radius of four in each
dimension:
\begin{lstlisting}
#pragma omp parallel for
for(int k=4; k < N-4; k++) {
 for(int j=4; j < N-4; j++) {
  for(int i=4; i < N-4; i++) {
   float lap = c0 * %V%[k][j][i]
  + c1 * ( %V%[ k ][ j ][i+1]+ %V%[ k ][ j ][i-1])
  + c1 * ( %V%[ k ][j+1][ i ]+ %V%[ k ][j-1][ i ])
  + c1 * ( %V%[k+1][ j ][ i ]+ %V%[k-1][ j ][ i ])
  + c2 * ( %V%[ k ][ j ][i+2]+ %V%[ k ][ j ][i-2])
  + c2 * ( %V%[ k ][j+2][ i ]+ %V%[ k ][j-2][ i ])
  + c2 * ( %V%[k+2][ j ][ i ]+ %V%[k-2][ j ][ i ])
  + c3 * ( %V%[ k ][ j ][i+3]+ %V%[ k ][ j ][i-3])
  + c3 * ( %V%[ k ][j+3][ i ]+ %V%[ k ][j-3][ i ])
  + c3 * ( %V%[k+3][ j ][ i ]+ %V%[k-3][ j ][ i ])
  + c4 * ( %V%[ k ][ j ][i+4]+ %V%[ k ][ j ][i-4])
  + c4 * ( %V%[ k ][j+4][ i ]+ %V%[ k ][j-4][ i ])
  + c4 * ( %V%[k+4][ j ][ i ]+ %V%[k-4][ j ][ i ]);
   U[k][j][i] = 2.f * %V%[k][j][i]- U[k][j][i] 
              + ROC[k][j][i] * lap;
}}}
\end{lstlisting}
The only array that is relevant for layer conditions here
is \verb.V.; the others (\verb.U. and \verb.ROC.)
are accessed in linear order without potential for
cache re-use.

\subsection{Analysis and \ecmm}

The IACA tool reports an overlapping core time of
$T_\mathrm{OL}=68\,\cycles$, which is mainly due to adds and frontend
stalls, while the load port limitation is at $T_\mathrm{nOL}=62\,\cycles$
(there are only 27 visible loads in the loop body, but
due to some integer register spilling four more cycles
are required).
Since the stencil radius is 4 in all directions, 
the block size in the $j$ direction must be chosen
so that the cache can hold nine layers. The $n$-thread layer condition is 
\bq 
9\cdot N\cdot b_j\cdot n\cdot 4\,\bytes <\frac{C_3}{2}\eos 
\eq 
In this case, four data streams per
thread will hit the memory and the code balance will be
$B_\mathrm{c}^\mathrm{mem}=16\,\bytes/\LUP$. If the layer condition
is only satisfied in L3, this cache will be hit by twelve streams per
thread, which already indicates that the transfer time is significant for
this kernel and that the main contribution to it does not come from
main memory. Indeed the \ecmm\ is \ecm{68}{62}{24}{24}{17}{\cycles},
which leads to a cycle prediction of \ecmp{68}{86}{110}{127}{\cycles}.
Hence, only $17/127\approx 13\%$ of the execution time
is attributed to $T_\mathrm{L3Mem}$, and the code will just barely
saturate at eight cores. If the layer condition is not satisfied at all,
the memory code balance will rise to $B_\mathrm{c}^\mathrm{mem}=48\,\bytes/\LUP$.
\begin{figure}[tb]
\centerline{\includegraphics*[width=0.9\linewidth]{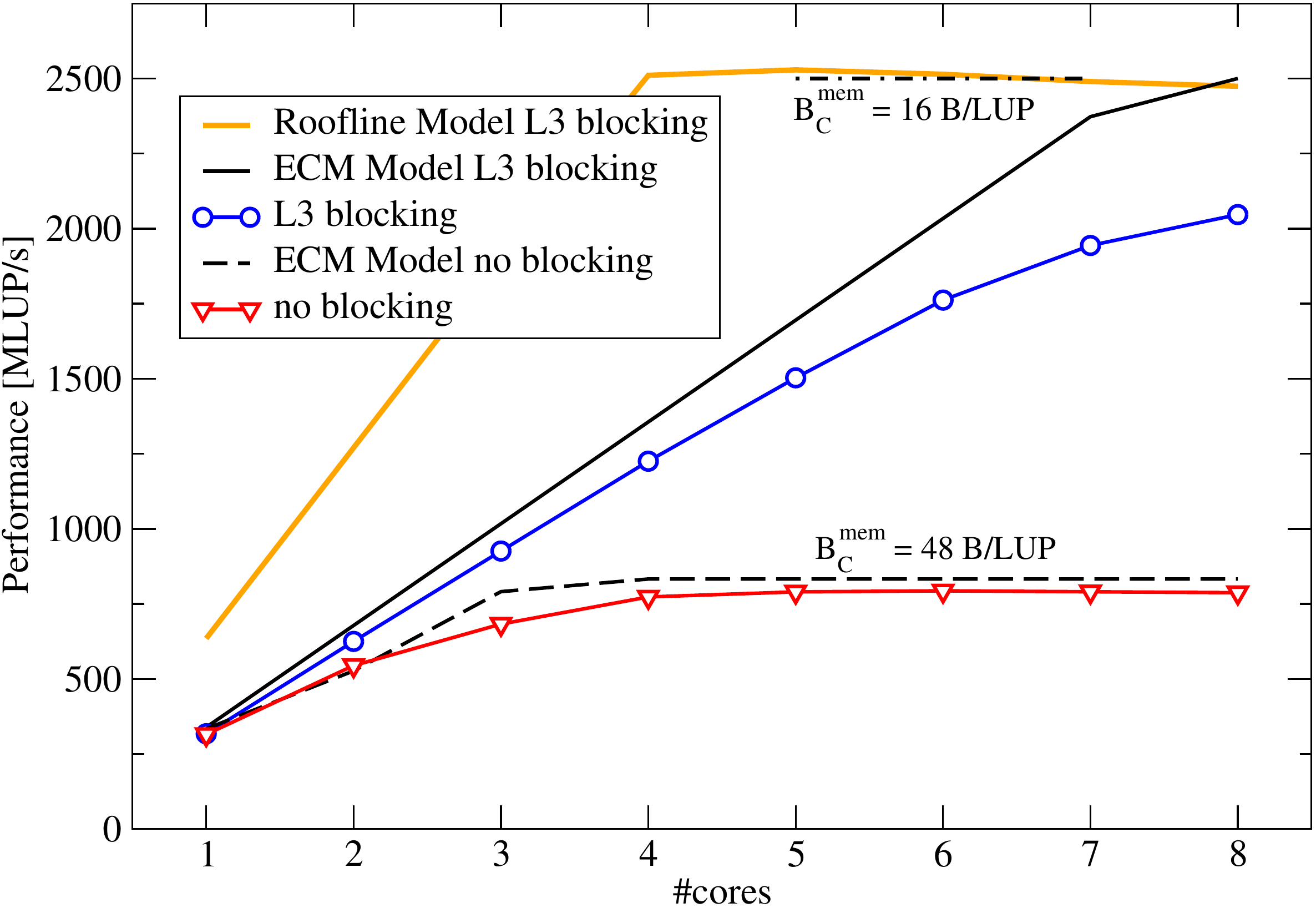}}
\caption{\label{fig:kaust-scaling}Performance scaling and
  \ecmm\ predictions of the 3D long-range stencil benchmark with (circles)
  and without (triangles) L3 blocking at a problem size of
  $480^3$ grid points across the cores of one SNB socket
  in single precision. The \rlm\ prediction is shown for reference.}
\end{figure}
In Fig.~\ref{fig:kaust-scaling} we show performance data on a SNB
socket with and without blocking. At the chosen problem size of
$480^3$, the layer condition is fulfilled without blocking for a
single thread, which is why the single-core performance and prediction
is the same in both cases. At two cores and above, however, 
blocking is required. The blocked code scales well almost
up to the full socket, as predicted by the \ecmm.

This stencil provides a good example for a situation where the
\rlm\ fails to deliver a useful performance estimate. Due to the large
in-core execution time, the loop is predicted to be core-bound until
the memory bandwidth is saturated, which happens at four cores (top
data set in Fig.~\ref{fig:kaust-scaling}). However, the other data
transfer contributions are far from negligible, so the \rlm\ is much
too optimistic here due to its overlapping assumption.

\subsection{Optimization opportunities}

The blocking optimization has almost removed the memory bottleneck,
and the \ecmm\ predicts that temporal blocking, which would eliminate
$T_\mathrm{L3Mem}$ altogether, would only have a minor effect on the
single core \emph{and the full socket}. In order to get any
performance improvement one would have to reduce the major
contribution in $T_\mathrm{mem}^\mathrm{ECM}$, which is
$T_\mathrm{nOL}$. If all core contributions including 
$T_\mathrm{nOL}$ could be shrunk by
50\%, the \ecmm\ would be \ecm{34}{31}{24}{24}{17}{\cycles}
and the prediction would be \ecmp{34}{55}{79}{96}{\cycles}.
The single-core code would be accelerated by 33\% and saturation
would set in at six cores, making temporal blocking a viable
optimization. Possible code transformations leading to 
a substantial improvement of $T_\mathrm{core}$ would have
to save on arithmetic operations as well as on loads. The
semi-stencil algorithm~\cite{delaCruz:2014} is one recent
example.

\section{Summary}
\label{sec:conclusion}
%


In this work we have reviewed the ECM performance model using simple
streaming kernels and refined it to clearly state its assumptions
regarding the overlap of different contributions to the single-core
runtime of a loop kernel across the memory hierarchy. By introducing a
short-hand notation we could ease the presentation of the model
results and performance predictions. We have
then applied the model to three stencil algorithms with different code
characteristics on an Intel SandyBridge processor. Using a 2D Jacobi
stencil as the initial example we were able to explain the
single-core performance vs.\ problem size by accurately calculating
the runtime contributions. The
performance and scalability impact of spatial blocking for
establishing ``layer conditions'' in different cache levels was
quantified using the \ecmm, and it was shown why
outer loop blocking is required although the layer condition does not
depend on the outer problem dimension. The performance of a more
complex 3D stencil with an expensive divide operation in the inner
loop was shown to be independent of the presence of the divide because
of the dominance of transfer times. Temporal blocking was predicted to
show a minor speedup on the single core but a major performance boost
on the full chip even if the divide operation is left untouched. Finally we
have used the model to show that a 3D long-range stencil cannot
saturate the memory bandwidth of the CPU at minimum in-memory code
balance because of in-cache contributions to the single-core
runtime. Any optimization attempt would thus have to aim at reducing
the in-cache runtime before more advanced techniques such as temporal
blocking are applied. 

It must be stressed that we have barely scratched the
surface of the vast range of known stencil optimizations, but 
the basic line of thinking has been 
clearly demonstrated. We have also shown when and why \rlm\
predictions fail to coincide with the \ecmm.
The \ecmm\ is to our knowledge the only model that can usefully
estimate the contributions to the single-core execution time of
streaming loop kernels, of which stencil codes are just one
example. It leads to a clear understanding of relevant bottlenecks and
potential optimization approaches but can mostly be set up using
``pencil and paper.'' Work is ongoing to build a simple
tool that can construct the model from a high-level description of the
code and the architecture, making it simpler to apply for the
non-expert.

One basic limitation of the \ecmm\ is its
``streaming assumption:'' The model is over-optimistic if the data transfers
through the memory hierarchy are latency-bound, which is the case,
e.g., in ``pointer chasing'' scenarios, or on processors for which
hardware and/or software prefetch mechanisms cannot hide the latency
cost. In addition, we do not know
yet how to incorporate the cost of non-temporal stores. Finally 
we have observed the model to be too optimistic for the memory data 
transfer time of tight loops with low $T_\mathrm{core}$
on processors with high-frequency memory interfaces.
These shortcomings may be ``fixed'' by introducing additional latency
penalties, but we consider such an approach undesirable, since it would
introduce additional parameters just for making the predictions more
accurate. The main goal of the model is, however, to understand trends
and bottlenecks, for which a precise performance prediction is often not
required.


\section*{Acknowledgment}

We thank Andrey Semin (Intel Germany) for useful discussions, Olaf
Schenk (USI Lugano) for providing the ``uxx'' benchmark case, and
Hatem Ltaief (KAUST) for providing the 3D long-range stencil case.
Part of this work was supported by the DFG priority programme 1648
``SPPEXA'' under the project ``EXASTEEL.''

\bibliographystyle{IEEEtran}
\bibliography{rrze,stencil}

\end{document}